\documentclass[12pt]{article}
\pdfmapfile{+euler.map}
\usepackage{blindtext}
\usepackage{graphicx} 
\usepackage{amsmath}
\usepackage{amsfonts}
\usepackage{calc}
\usepackage{amssymb}
\usepackage{amsthm}
\usepackage[margin=1in,footskip=0.25in]{geometry}
\usepackage{pict2e}
\usepackage{color}
\newcommand{\RefFrac}[2]{{\dfrac{#1}{#2}}}
\usepackage{relsize}
\usepackage{verbatim}

\usepackage{accents}

\makeatletter
\DeclareRobustCommand{\loplus}{\mathbin{\mathpalette\dog@lsemi{+}}}
\DeclareRobustCommand{\lotimes}{\mathbin{\mathpalette\dog@lsemi{\times}}}
\DeclareRobustCommand{\roplus}{\mathbin{\mathpalette\dog@rsemi{+}}}
\DeclareRobustCommand{\rotimes}{\mathbin{\mathpalette\dog@rsemi{\times}}}

\newcommand{\dog@rsemi}[2]{\dog@semi{#1}{#2}{-90,90}}
\newcommand{\dog@lsemi}[2]{\dog@semi{#1}{#2}{270,90}}
\newcommand{\dog@semi}[3]{%
  \begingroup
  \sbox\z@{$\m@th#1#2$}%
  \setlength{\unitlength}{\dimexpr\ht\z@+\dp\z@\relax}%
  \makebox[\wd\z@]{\raisebox{-\dp\z@}{%
    \begin{picture}(1,1)
    \linethickness{\variable@rule{#1}}
    \roundcap
    \put(0.5,0.5){\makebox(0,0){\raisebox{\dp\z@}{$\m@th#1#2$}}}
    \put(0.5,0.5){\arc[#3]{0.5}}
    \end{picture}%
  }}%
  \endgroup
}
\newcommand{\variable@rule}[1]{%
  \fontdimen8  
  \ifx#1\displaystyle\textfont3\else
    \ifx#1\textstyle\textfont3\else
      \ifx#1\scriptstyle\scriptfont3\else
        \scriptscriptfont3\relax
  \fi\fi\fi
}
\makeatother

\theoremstyle{definition}
\newtheorem{tw}{Theorem}[section]

\newtheorem{defi}[tw]{Definition}
\newtheorem*{dow}{Proof}
\newtheorem{uw}[tw]{Remark}

\newtheorem{lem}[tw]{Lemma}
\newtheorem{prop}[tw]{Proposition}

\newtheorem{col}[tw]{Corollary}

\newtheorem*{tw*}{Theorem 1}

\newtheorem{exam}[tw]{Example}

\DeclareMathOperator{\diverg}{div}
\DeclareMathOperator{\crl}{curl}
\DeclareMathOperator{\codim}{codim}
\DeclareMathOperator{\der}{Der}
\DeclareMathOperator{\im}{im}
\DeclareMathOperator{\nos}{supp}
\DeclareMathOperator{\card}{card}
\DeclareMathOperator{\spam}{span}
\DeclareMathOperator{\rak}{rank}

\title{ Nonlinear wave superpositions and quasi-rectifiable Lie modules}
\author{Łukasz Chomienia$^{1}$,
\text{Alfred Michel Grundland*}$^{2,3}$\\
\textsuperscript{1}Department of Mathematics and Statistics, University of Jyväskylä,\\ P.O. Box 35 (MaD), FI-40014, Jyväskylä, Finland\\
e-mail : lukasz.l.chomienia@jyu.fi\\
\textsuperscript{2}Centre de Recherches Math{\'e}matiques, Universit{\'e} de Montr{\'e}al,\\ Succ. Centre-Ville, CP6128, Montr{\'e}al (QC) H3C 3J7, Canada\\
\textsuperscript{3}D{\'e}partement de Math{\'e}matiques et d'Informatique, Universit{\'e} du Qu{\'e}bec\\
à Trois-Rivières, CP 500, Trois-Rivières (QC) G9A 5H7, Canada
\\
e-mail : alfred.michel.grundland@umontreal.ca}

\date{\today}

\numberwithin{equation}{section}
\begin{document}
\maketitle


\begin{abstract}

This paper presents a study of nonlinear superpositions of Riemann wave solutions admitted by  hyperbolic first-order systems. We focus on the Euler system and non-elastic wave superpositions that cannot be decomposed into pairwise independent interactions of waves. We use the property of quasi-rectifiability of the families of vector fields, which  imposes certain conditions on the commutators of these vector fields. They enable us to find a parametrization of the region of superpositions of Riemann waves which leads to a simplification of the initial system. In order to identify non-elastic superpositions we prove that a class of  associated Lie modules can be uniquely transformed into a real Lie algebra through an angle-preserving transformation. We  select a particular basis of vector fields associated with a given module which ensures the property of quasi-rectifiability. That, in turn, allows us to construct the reduced form of the Euler system for which a non-elastic superposition of two Riemann waves is then derived. A study of the geometry of the manifold of  non-elastic wave superpositions in terms of deformations of submanifolds corresponding to the Lie algebras is performed. Finally, we adapt the described approach to the general form of a hydrodynamic-type system, i.e. to arbitrary Lie modules of vector fields, providing the criteria for their quasi-rectifiability. A geometric interpretation of non-elastic wave superpositions in this system is given.
\end{abstract}

\bigskip
* corresponding author
\bigskip

\newpage
\bigskip


\tableofcontents

\section{Introduction}
\text{}\\
\indent In this paper, we analyze the evolution of Riemann waves that enter in the nonlinear superpositions obtained from quasilinear homogeneous hyperbolic first-order systems of partial differential equations (PDEs) in two independent variables. This subject has been studied mainly for  elastic wave superpositions in the context of determining particular non-stationary fluid flows (see, e.g., \cite{Jef, Maj, Pera, Whi}).  The  cases of the non-elastic Riemann wave superpositions have been treated mostly by numerical methods, and no general analytical approach to this problem  exists up to now. This could be due, in part, to the fact that the problem of Riemann wave superpositions has traditionally been considered in the framework of the method of characteristics, which is computationally difficult and produces results in an implicit form.
In this paper, we apply a different approach to this problem, using the apparatus of  modern Lie algebra theory, which provides tools, such as Lie modules of vector fields, that are especially suitable for the description of wave superposition phenomena. This approach has been recently developed in \cite{Gru3} for elastic Riemann wave superpositions. Its main element is the introduction of the notion of  quasi-rectifiability which defines certain properties of the vector fields, ensuring the existence of integrable surfaces and enabling their construction. This approach cannot be applied directly to the case of non-elastic wave superpositions and requires an adaptation, which is performed in this paper.

     A specific feature of the method proposed in this paper is to assume that $C^\infty$-Lie modules of families of vector fields can be identified  with infinite-dimensional Lie algebras. This fact significantly   facilitates the analysis of non-elastic cases of wave superpositions. We develop this approach for  the Euler system in (1+1)-dimensions.
      We prove that for the Euler system the smallest Lie algebra corresponding to the case of a non-elastic wave superposition has the structure of a semi-direct sum of an infinite-dimensional Lie algebra and a certain finite-dimensional one. We present two decompositions of this kind which depend on the choice of an Abelian ideal. Since the infinite-dimensional component is Abelian, the non-elastic wave superpositions in this system are represented by the finite-dimensional algebra.
We submit this algebra to the angle-preserving transformation which provides a new basis for the associated vector fields, such that they become quasi-rectifiable. This property ensures that, after an appropriate rescaling of these vector fields, we can apply the Frobenius theorem and determine the parametrization of the region of non-elastic wave superpositions. This parametrization provides a reduced form of the Euler system, which exclusively admits these types of superpositions and can be solved analytically as illustrated by the included example. It also enables  a description of the manifold structures representing non-elastic wave superpositions in terms of surfaces spanned by the quasi-rectifiable families of vector fields. Moreover, we show that after certain modifications, these structures yield a parallel transport of these surfaces.

    The described approach can be extended to arbitrary hydrodynamic-type systems. To this end, we introduce a transformation that maps the class of quasi-rectifiable Lie modules associated with such systems to real Lie algebras in a way that preserves angles between the vector fields forming a given Lie module. In this way, we associate each module with the unique (up to an isomorphism) real Lie algebra and the corresponding Lie group. For the connected Lie groups associated with the interacting waves, we make use of the Nomizu theorem \cite{Nom} concerning the existence of a natural affine connection - one in which one-parameter subgroups serve as geodesics. An analysis of various classes of decomposable Lie algebras associated with waves provides a geometric interpretation of the process of non-elastic wave superpositions.

The paper is organized as follows. {In \textbf{Section 2}, we} present a summary of results concerning the quasi-rectifiability of families of vector fields that are used to construct integral manifolds and some basic notions associated with nonlinear Riemann wave superpositions. This section is included in order to make the paper self-contained and avoid back-and-forth between the paper and associated references. {In \textbf{Section 3}, we} discuss the difference between  the elastic and non-elastic Riemann wave superpositions and their connection to the property of quasi-rectifiability of vector fields. \textbf{Section 4} gives a brief account of certain geometric aspects of the quasi-rectifiability of vector fields defined in three-dimensional space. \textbf{Section 5} is devoted to the study of wave interactions in the one-dimensional  Euler system. After briefly reviewing known results on the wave analysis of this system, we describe the Lie module of waves associated with the Euler system in terms of real Lie algebras. We then show how the finite-dimensional component of this Lie algebra can be derived via rescaling. Building on these results, we offer a geometric description of the manifold of wave superpositions and discuss deformations leading to a description via parallel transport. Finally, we show how these findings yield a reduced form of the Euler system. In \textbf{Section 6}, we generalize the analysis carried out for the Euler system to arbitrary Lie modules associated with hydrodynamic-type systems. We prove that such modules can be rescaled to yield finite-dimensional real Lie algebras.  {In \textbf{Appendix A}, we} discuss the rescaling of non-quasi-rectifiable Lie algebras and investigate the Lie group associated with the Euler system. {In \textbf{Appendix B}, we} construct a particular class of explicit solutions of the reduced Euler system.

\section{Preliminaries}

        Consider a homogeneous strongly hyperbolic quasilinear system of first-order PDEs  written in the form
		\begin{equation}\label{syst_PDEs_2}
			{u_t+A(u)u_{x^1}=0}
		\end{equation}
		{in two independent variables $x=\left(t=x^0,x^1\right)\in\mathbb{R}^2$ (below, $x^1$ is also denoted by $x$ when paired with $t$)} and $q$ dependent variables $u=\left(u^1(x),...,u^q(x)\right)\in\mathbb{R}^q$,
        {where $A=(A^\alpha_\beta)$ is a $q\times q$ matrix-valued function of $u$.} All functions are assumed to be smooth. It suffices to consider the solutions defined on a neighborhood of $x=0$. The system (\ref{syst_PDEs_2}) is written in evolutionary form and is assumed to be subjected to differential constraints (DCs). All first-order partial derivatives of $u^{\alpha}$, $\alpha=1,...,q$ are decomposable		\begin{equation}\label{deriv_u_decomp_2}
        \begin{split}
			&\dfrac{\partial u^\alpha}{\partial x^i}=\sum_{s=1}^{k}\xi^s(x)\gamma_s^{\alpha}(u)\lambda_i^s(u),\quad\alpha=1,\dots,q,\quad i=0,1,\\
            &\gamma_s=\sum_{\alpha=1}^{q}\gamma_s^{\alpha}(u)\dfrac{\partial}{\partial u^{\alpha}}\in T_u\mathbb{R}^q\quad\mbox{and}\quad\lambda^s=v_s(u)dt-dx^1{\in(\mathbb{R}^{2})^*},\quad s=1,...,k
        \end{split}
		\end{equation}
        {where $i=0,1$ labels the independent variables, $s=1,\dots,k$ labels the wave families, and $\xi^s(x)$ are nonvanishing scalar functions.} The quantities $v_1(u)<...<v_k(u)$ are the eigenvalues of the matrix $A=(A^{\alpha}_{\beta})$. The vector fields $\gamma_1,...,\gamma_k$ are assumed to be linearly independent at any point on $\mathbb{R}^q$
        \begin{displaymath}
\gamma_1\wedge ...\wedge\gamma_k\neq 0.
        \end{displaymath}
		The pairs of functions $\left(\gamma_s,\lambda^s\right)$ are assumed to satisfy the algebraic eigenvalue conditions
		\begin{equation}\label{algebra-ic_2}
			\sum_{\beta=1}^{q}\big(v_s\delta_\beta^\alpha-A_\beta^{\alpha}(u)\big)\gamma_s^\beta=0,\quad\det\big(v_s\delta_\beta^\alpha-A_\beta^{\alpha}(u)\big)=0, \quad s=1,\dots,k,\quad\alpha=1,\dots,q,
		\end{equation}
         The functions $\left(u^1,\dots,u^q,\xi^1,\dots,\xi^k\right)$ are considered to be unknown functions of $x$ of the system (\ref{syst_PDEs_2}) and (\ref{deriv_u_decomp_2}). {System \eqref{syst_PDEs_2} is of hydrodynamic type; the equations \eqref{deriv_u_decomp_2} impose additional differential constraints on its solutions.}

The compatibility conditions for the DCs (\ref{deriv_u_decomp_2}), with arbitrary functions $\xi^s(x)$, require \cite{Pera} that {the commutator $[\gamma_{{s}},\gamma_{{l}}]$ of any two vector fields $\gamma_{{s}}$ and $\gamma_{{l}}$ lie in the span of these fields}, i.e., 
         \begin{equation}\label{conditions_2}
             \left[\gamma_{{s}},\gamma_{{l}}\right]\in \text{span}\left\{\gamma_{{s}},\gamma_{{l}}\right\},\quad {s\neq l}\in\{1,\dots,k\},
         \end{equation}
         where 
         \begin{equation}
            \left[\gamma_{{s}},\gamma_{{l}}\right]=\gamma_{{s}}^\alpha\dfrac{\partial}{\partial u^\alpha}\gamma_{{l}}-\gamma_{{l}}^\alpha\dfrac{\partial}{\partial u^\alpha}\gamma_{{s}}.
         \end{equation}
         Conditions (\ref{deriv_u_decomp_2}) and (\ref{conditions_2}) are a starting point to formulate the problem of superposition of Riemann waves (Section 3).

\subsection{Riemann wave solutions}
 We begin by discussing  certain classes of solutions of (\ref{syst_PDEs_2}) according to the rank of the matrix of  {derivatives of $u$,} i.e., $\operatorname{rank} \left|\frac{\partial u^\alpha}{\partial x^i}\right|=k$. For example, a solution $u(x)$ has a rank equal to zero if and only if it is constant. A special case of a rank-$1$ solution is a Riemann wave solution of (\ref{syst_PDEs_2}).
         
         {Let a pair $(\gamma,\lambda)$ satisfy \eqref{algebra-ic_2}} (called a simple element) and a map $f\colon\mathbb{R}\to \mathbb{R}^q$ be an integral curve $\Gamma$ of the vector field $\gamma^{\alpha}(u)\frac{\partial}{\partial u^{\alpha}}$ {parametrized by $r,$} i.e.,
         \begin{equation}
             \Gamma \colon \; \;\dfrac{df^\alpha}{dr}=\gamma^\alpha\left(f^1(r),\dots,f^q(r)\right).
         \end{equation}
         {For a fixed wave family $s\in\{1,\dots,k\}$, take $(\gamma,\lambda)=(\gamma_s,\lambda^s)$. The coefficients of $\lambda^s$ along $\Gamma$ are obtained by composition with $f$; in particular, we write $v_s(r):=v_s(f(r))$.}
         Then, the implicitly defined relations between the variables $u,r$ and $x$ given by 
         \begin{equation}
             u^\alpha=f^\alpha(r),\quad r=\varphi\left(v_s(r)t-x^1\right),\quad \text{for every} \quad s=1,...,k,
             \label{Riemann wave}
         \end{equation}
         {where $\varphi\colon\mathbb{R}\to\mathbb{R}$ is an arbitrary differentiable function defining the wave profile, constitute} an exact simple Riemann wave solution of system \eqref{syst_PDEs_2}. The scalar function $r(x)$ associated with the wave vector $\lambda$ is called a Riemann invariant.

         The differential of the function $u^{\alpha}$ following from the expression \eqref{Riemann wave} is decomposable in terms of the vectors $\gamma$ and $\lambda,$ and satisfies the initial system \eqref{syst_PDEs_2}, and the differential constraints \eqref{deriv_u_decomp_2}. We have
         \begin{equation}
             du=\frac{d f}{dr}dr=\frac{\varphi'(z_s)}{1-\varphi'(z_s) \frac{d v_s}{dr}t} \gamma_s\otimes\lambda^s,  \quad \varphi'(z_s)=\frac{d\varphi}{dz_s}, \quad z_s=v_s(r)t-x^1,\quad s=1,...,k.
         \end{equation}

         Note that the Riemann wave constitutes a generalization of the plane harmonic wave 
         
         \begin{equation}
         u= \operatorname{Re} \left( \gamma e^{i(\omega t-{\Vec{\ell}}\cdot \Vec{x})} \right), \quad \lambda= \left( \omega, {\Vec{\ell}}\right), \nonumber \end{equation}

         \noindent to the case of a wave with an arbitrary profile. This case was first formulated in 1808 by S.D. Poisson \cite{Poisson} who studied a nonlinear wave propagating in an isothermal gas governed by the equation $u_t+u u_x=0$. Its solution takes the implicit form 

         \begin{equation}
             u=F\left( x-(u+c) t\right),\qquad c=\left(\frac{\kappa T}{m} \right)^\frac{1}{2}, 
             \end{equation}
where $F$ is an arbitrary differentiable function, $c$ is the velocity of gas flow, $\kappa$ is the adiabatic exponent, and $T$ and $m$ are the temperature and mass of the element of gas. The theory of wave propagation and interactions was further developed by B. Riemann who introduced new tools to this analysis. He investigated \cite{Rie,Rie2} a nonstationary isentropic flow  of an ideal compressible fluid in one spatial dimension with density $p=A \rho^\kappa$. He demonstrated that two waves propagating in opposite directions after superposition could be separated again, resulting in two single waves of the same type as those imposed in the initial data.

\subsection{Quasi-rectifiable families of vector fields}
    
   The methodological approach used in this paper is based on the notion of quasi-rectifiability of vector fields, which was developed in \cite{Gru3}. Below we summarize the main elements of this approach. 
   \begin{tw} \label{straightening}
        (Straightening of vector fields \cite{Gru3})
Let $X_1,\dots,X_r$ be a family of vector fields defined on an $n$-dimensional manifold $N$ such that 
\begin{equation}
           X_1 \wedge \dots \wedge X_r \neq 0  \quad \text{ at any point on } N.
       \end{equation}
There exists a local coordinate system ${x^1,\dots, x^n }$ on $N$ such that the first integrals of each vector field $X_i$ are given by the equations

\begin{equation}
    X_i x^j =0, \quad \text{for }\quad i\neq j,\quad  i=1,\dots,r, \quad j=1,\dots, n,
\end{equation}

\noindent if and only if the Lie bracket of any pair of vector fields $X_i$ and $X_j$ is a linear combination of $X_i$ and $X_j$ with coefficients which are not necessarily constant, i.e.,

\begin{equation}
    \left[X_i, X_j\right]=f^i_{ij}X_i+f^j_{ij}X_j, \quad 1\leq i<j \leq r <n \quad \text{(no summation)},
\end{equation}

{\noindent where the $r(r-1)$ coefficient functions $f_{ij}^i$ and $f^j_{ij}$, $1\leq i<j\leq r$, belong to $C^\infty(N)$.}\qed
       
   \end{tw}

  \begin{defi}\label{qfam} {\text{(Quasi-rectifiable family of vector fields} \cite{Gru3})}
  A family of vector fields $X_1, \dots, X_r$ defined on a manifold $N$ {is said to be \emph{quasi-rectifiable} if there exists} a local coordinate system $\left\{x^1,\dots x^n \right\}$ on $N$ such that each vector field $X_j$ has the form

 \begin{align}
    \begin{aligned}
        &X_j=g_j(x^1,\dots,x^n)\frac{\partial}{\partial x^{j}}, \quad  j \in \{1,\dots, r \} \quad \text{(no summation)},\\ 
    \quad & X_1 \wedge \dots \wedge X_r \neq 0 \quad \text{at any point on $N$},
    \label{quasi}
    \end{aligned}
\end{align}
for some functions $g_1,\dots, g_r \in C^\infty(N)$. {A family that does not satisfy this condition is called non-quasi-rectifiable.} The coordinate expression \eqref{quasi} is called a quasi-rectifiable form for the vector fields $X_1,\dots, X_r$ on $N.$ 

\end{defi}

   {We call a Lie module quasi-rectifiable with respect to a specified basis if that basis is a quasi-rectifiable family. A real Lie algebra of vector fields is called quasi-rectifiable if it admits a basis of this kind. The quasi-rectifiability of a specified family is basis-dependent.}

\begin{tw}\label{tw2.3}
     (Rescaling of vector fields \cite{Gru3}) \label{modified}
     Let $X_1,\dots, X_r$ be a quasi-rectifiable family of vector fields defined on an $n$-dimensional manifold $N$ such that
\begin{align}
        \begin{aligned}
            & \left[X_i, X_j\right]=f^i_{ij}X_i+f^j_{ij}X_j, \quad 1\leq i<j \leq r <n \quad \text{(no summation)},\\
        \quad & X_1 \wedge\dots \wedge X_r \neq 0\quad \text{at any point on $N.$}
        \end{aligned}
    \end{align}
Let $D$ be a distribution spanned by the vector fields $X_1,\dots, X_r$ and let $\eta^1,\dots, \eta^r$ be a collection of one-forms that is dual to the collection of the vector fields $X_1,\dots, X_r$ on $N$. The contraction of the one-forms $\eta^i$ and the vector fields $X_j$ is 
\begin{equation}
    \eta^i \left(X_j\right)=\delta^i_j, \quad i,j \in \left\{1,\dots,r\right\}. 
\end{equation}

\noindent Let the nonvanishing functions $h_1,\dots, h_r \in C^\infty (N)$ be the rescaling functions for $X_1,\dots, X_r$. {Then the rescaled vector fields $h_1X_1,\dots,h_rX_r$ commute if and only if the dual one-forms $h_1^{-1}\eta^1,\dots,h_r^{-1}\eta^r$ are closed on each integral leaf of $D$, and hence locally exact on that leaf of  $D$, i.e., } 

\begin{equation}
    {d(h_i^{-1}\eta^i)\big|_D=0} , \quad  i \in \left\{1,\dots,r\right\}.
    \label{8}
    \end{equation}
    \qed
\end{tw}  
\indent {Locally on each integral leaf of $D$, choose coordinates $x^1,\dots,x^r$ such that $dx^i=h_i^{-1}\eta^i|_D$. Then $h_iX_i=\frac{\partial}{\partial x^i},$ and thus each $X_i$, $i\in\{1,\dots,r\}$, has the quasi-rectifiable form \eqref{quasi} in the sense of Definition \ref{qfam}.}
\subsection{Representation of $C^\infty$-Lie module via infinite-dimensional Lie algebras}\label{infinite-dimensional Lie algebras}
The family of vector fields \eqref{subeq} corresponding to the Euler system (\ref{euler}) constitutes a Lie module $\left\{|X_+,X_0,X_-|\right\}$ over $C^{\infty}$ since these vector fields are closed with respect to the Lie bracket:

\begin{equation}\label{eulercom2}
    \begin{aligned}
      & \left[X_+,X_-\right]=\dfrac{1-k}{2}X_++\dfrac{k-1}{2}X_-, \quad
      \left[X_+,X_0\right]=\dfrac{1}{4\rho}X_+-\dfrac{1}{4\rho}X_--X_0,\\
      &\left[X_0,X_-\right]=\dfrac{1}{4\rho}X_+-\dfrac{1}{4\rho}X_-+X_0.
    \end{aligned}
\end{equation}

Note that any set of vector fields closed with respect to a Lie bracket can be interpreted in a twofold way. 
\begin{lem}
Let $V$ be a $C^{\infty}$-Lie module generated by a finite family of smooth vector fields. Then $V$ generates a unique infinite-dimensional real Lie algebra. \qed
\end{lem}

         We now proceed to determine the structure of the infinite-dimensional real Lie algebra generated by the vector fields $X_-,X_0,X_+.$ Let us first recall the following definition:
         
         \begin{defi}
         Let $W_{\geqslant -1}$ be an infinite-dimensional real algebra with a basis $\left\{a_n\right\}_{n \in \mathbb{N}\cup \left\{0\right\}}$ equipped with the bracket $$\left[a_n,a_m\right]:=\left(n-m\right)a_{n+m}.$$ The algebra $W_{\geqslant -1}$ is called the one-sided Witt algebra. 
         \end{defi}

         \begin{uw}
         The Witt algebra is sometimes introduced as the Virasoro algebra without central extension \cite{Hua}.
         Note that there is only one non-equivalent realization of the Witt algebra over $\mathbb{R}$-space \cite{Hua}, thus these two definitions coincide.
        \end{uw}
         

        \begin{uw}
            Despite the fact that the category of Lie algebras is not an Abelian category, it possesses initial objects, kernels and co-kernels. This means that the First Isomorphism Theorem applies, i.e., if $\phi\colon A \to B$ is a homomorphism, then the isomorphism ${A/\ker \phi \simeq \im \phi,}$ holds. Moreover, the Second Isomorphism Theorem and the Third Isomorphism Theorem apply as well.
        \end{uw}
         Note that the real Witt algebra $W_{\geqslant -1}$ can be realized as an infinite-dimensional Lie algebra of polynomials in one real variable $\mathbb{R}[z]$ with the Lie bracket $$\left[z^n,z^m\right]:=\left(n-m\right)z^{n+m},\quad n,m \in \mathbb{N}\cup \left\{0\right\}.$$ 
         \begin{defi}
         Let $I_1 = \mathbb{R}\left[y\right]$ and $I_2 = \mathbb{R}\left[x\right]$ be real Lie algebras equipped with the zero Lie bracket, i.e., $\left[\cdot_1,\cdot_2\right]\equiv 0.$
By a shift operator $\Psi\colon W_{\geqslant -1} \to \der I_1$ we mean a linear map defined by 
\begin{equation}\label{shift}
\Psi_{z^n}\left(y^m\right):= -\left(m+1\right)y^{m+n}, \quad n,m \in \mathbb{N} \cup \left\{0\right\},
\end{equation}
where $\der I_1$ denotes the set of derivations of the algebra $I_1,$ i.e., the set of all homomorphisms of $I_1$ satisfying the Leibniz rule with respect to the Lie bracket. By analogy, we define a shift operator $\Phi$ taking values in $\der I_2.$ 
\end{defi}

The following theorem allows for the identification of the infinite-dimensional real Lie algebra generated by the vector fields $X_i,\; i\in \left\{+,-,0\right\}$ with a certain finite-dimensional one.

\begin{tw}\label{infimp}
    The smallest (with respect to inclusion) real Lie algebra containing the vector fields $X_+,\; X_-,\; X_0$ is isomorphic to  
    \begin{equation}\label{extra1}\mathcal{A} \simeq I^- \roplus_{\Theta} K^- ,\end{equation}
    where $I^-$ is an infinite-dimensional real Abelian Lie algebra, $\Theta$ is a shift operator, and $K^-$ is a three-dimensional real Lie algebra which are all direct sums of the unique $2$-dimensional non-Abelian real Lie algebra and the unique $1$-dimensional real Lie algebra. The Abelian ideal $I^-$ is given by 
    \begin{equation}\label{extra2}
    \begin{aligned}
    I^-:= \spam\left\{\rho^{-1}\left(X_+-X_-\right),\rho^{-2}\left(X_+-X_-\right),\rho^{-3}\left(X_+-X_-\right), \dots\right\},
    \end{aligned}
    \end{equation}
    
    \begin{dow}
        To shorten the computations we assume that $k=1$, but the case of an arbitrary constant $k$ can be considered in an analogous way. 

        We limit the following proof to the consideration of the decomposition $\mathcal{A} \simeq I^- \roplus_{\Theta} K^-.$ A proof of the alternative decomposition $\mathcal{H} \simeq I^+ \roplus_{\Theta} K^+$ follows a similar line. 
Let us define $Z:= X_+ - X_-.$ We recall that, for $k=1$, we obtain 
        \begin{equation}
            \begin{aligned}
                \left[X_+,X_-\right]=0,\quad 
                \left[X_+,X_0\right]=\frac{1}{4\rho}\left(X_+-X_-\right)-X_0 = \frac{1}{4\rho}Z - X_0,\\
                \left[X_-,X_0\right]=-\frac{1}{4\rho}\left(X_+-X_-\right)-X_0 = -\frac{1}{4\rho}Z-X_0.
            \end{aligned}
        \end{equation}
        Let us observe that 
        \begin{equation}
            \begin{aligned}
                \left[Z,X_0\right]=\frac{1}{2\rho}Z,\quad
                \left[Z,X_+\right]=\left[Z,X_-\right]=0.
            \end{aligned}
        \end{equation}
        By $\mathcal{A}$ we denote a real vector space generated by $X_0,X_-,X_+$ and $\rho^{-n}Z$ for $n\in \left\{1,2,3,\dots\right\},$ i.e.,
        $$\mathcal{A} := \spam \left\{X_+,X_-,X_0,\rho^{-1}Z, \rho^{-2}Z, \rho^{-3}Z,\dots\right\}.$$
        Direct computations show that 
        \begin{equation}
            \begin{aligned}
                \left[\rho^{-n}Z, \rho^{-m}Z\right]=0,\quad
                \left[X_0,\rho^{-n}Z\right]=-\left(n+\frac{1}{2}\right)\rho^{-\left(n+1\right)}Z,\\
                \left[X_+,\rho^{-n}Z\right]=-n\rho^{-n}Z,\quad
                \left[X_-,\rho^{-n}Z\right]=-n\rho^{-n}Z.
            \end{aligned}
        \end{equation}
        This shows that the Lie subalgebra 
        $I := \spam \left\{\rho^{-n}Z\right\}_{n\in \left\{1,2,3,\dots\right\}}$ is an ideal of the real Lie algebra $\mathcal{A}.$ 

        We denote the elements of the quotient Lie algebra $K:=\mathcal{A}/I$ as follows:
        $$\kappa_0 :=X_0 + I,\quad \kappa_- := X_- +I,\quad \kappa_+ := X_+ + I.$$ Then it is easy to check that the quotient Lie algebra 
        $$K = \left\{X_0+I, X_-+I,X_++I\right\}=\left\{\kappa_0,\kappa_-,\kappa_+\right\}$$ has the quotient Lie bracket relations:
        \begin{equation}
            \begin{aligned}
                \left[\kappa_-,\kappa_0\right]_K = -\frac{1}{4}\rho^{-1}Z-X_0=-\kappa_0,\quad
                \left[\kappa_+,\kappa_0\right]_K = \frac{1}{4}\rho^{-1}Z-X_0=-\kappa_0,\quad
                \left[\kappa_+,\kappa_-\right]=0.
            \end{aligned}
        \end{equation}
        By a proper change of basis of the Lie algebra $K$ it follows that $$K \simeq L(2,1)\oplus L(1,0),$$ where $L(2,1)$ stands for the unique real non-Abelian $2$-dimensional Lie algebra and $L(1,0)$ is the unique real $1$-dimensional Lie algebra.
For each element $\kappa_0, \kappa_-, \kappa_+$ we define the mappings $\Theta_{\kappa_i}\colon I \to I,\; i\in \left\{0,-,+\right\}$ as
        \begin{equation}
            \begin{aligned}
                &\Theta_{\kappa_+}\left(\rho^{-n}Z\right) := -n\rho^{-n}Z,\quad
                \Theta_{\kappa_-}\left(\rho^{-n}Z\right) := -n\rho^{-n}Z,\\
                &\Theta_{\kappa_0}\left(\rho^{-n}Z\right) := -\left(n+\frac{1}{2}\right)\rho^{-\left(n+1\right)}Z.
            \end{aligned}
        \end{equation}
        It is easy to check that each $\Theta_{\kappa_i},\; i\in \left\{0,-,+\right\}$ is a derivation of $I,$ i.e., it satisfies the Leibniz rule with respect to the Lie bracket: $$\Theta_{\kappa_i}\left(\left[\rho^{-n}Z,\rho^{-m}Z\right]\right) = \left[\Theta_{\kappa_i}\left(\rho^{-n}\right)Z,\rho^{-m}Z\right]+\left[\rho^{-n}Z,\Theta_{\kappa_i}\left(\rho^{-m}Z\right)\right].$$
        Next, by linearity, we uniquely extend the mapping $\Theta\colon K \to \der I,$ from the basis $\left\{\kappa_0, \kappa_-, \kappa_+\right\}$ to the whole Lie algebra $K.$ 

        Let us consider a direct sum $I \oplus K$ equipped with a bracket relation defined on the basis as
        \begin{equation}\label{swiastak}
            \begin{aligned}
                &\left[\rho^{-n}Z\oplus 0,0 \oplus \kappa_0\right]:=-\Theta_{\kappa_0}\left(\rho^{-n}w\right) \oplus 0 = -\left(n+\frac{1}{2}\right)\rho^{-\left(n+1\right)}Z \oplus 0,\\
                &\left[\rho^{-n}Z \oplus 0, 0 \oplus \kappa_+\right]:= -\Theta_{\kappa_+}\left(\rho^{-n}Z\right)\oplus 0 = -n\rho^{-n}Z \oplus 0,\\
                &\left[\rho^{-n}Z \oplus 0, 0 \oplus \kappa_-\right] := -\Theta_{\kappa_-}\left(\rho^{-n}Z\right)\oplus 0 = -n\rho^{-n}Z \oplus 0.
            \end{aligned}
        \end{equation}
        Applying the general formula for the semidirect sum of two Lie algebras, we find that the set $I \oplus K$ constitutes a Lie algebra which is the semidirect sum of $I$ and $K$ with respect to the mapping $\Theta,$ that is $I \roplus_{\Theta} K.$ 

        Finally, it is clear that a mapping $Q\colon  I \roplus_{\Theta} K \to \mathcal{A}$ given on the basis by
        \begin{equation}
            \begin{aligned}
                &Q\left(\rho^{-n}Z \oplus 0\right) := \rho^{-n}Z,\quad n\in \left\{1,2,3,\dots\right\},\\
                &Q\left(0\oplus \kappa_0\right) := \kappa_0,\quad
                Q\left(0 \oplus \kappa_+\right) := \kappa_+,\quad
                Q\left(0 \oplus \kappa_-\right):= \kappa_-,
            \end{aligned}
        \end{equation}
        is an isomorphism of the Lie algebras $I \roplus_{\Theta} K$ and $\mathcal{A}.$ 
        \qed
    \end{dow}
\end{tw}

\begin{uw}
    The decomposition (\ref{extra1}) for the algebra $\mathcal{A}$ (and analogously for $\mathcal{H}$) can be interpreted as follows.
    \begin{itemize}
        \item Due to the fact that $I^-\subset\mathcal{A}$ is an Abelian ideal, the whole qualitative behavior of interacting waves is encoded in $K^-$.
        \item The quantitative part of the wave interactions involving the varying density $\rho$ is encoded in the infinite-dimensional component $I^-$.
        \item The infinite-dimensional character of the Lie algebra $\mathcal{A}$ comes from the fact that it contains `higher order' iterations corresponding to sequences of wave superpositions.
    \end{itemize}
\end{uw}

\subsection{Rescaling a Lie module related to the Euler system}\label{euler-4.2}

A  transformation of the $C^{\infty}$-Lie module $\left\{|X_-,X_0,X_+|\right\}$ of eigenvectors of the Euler system \eqref{euler} to the finite-dimensional Lie algebra $K$ should preserve angles between each pair of vector fields and should be reversible. This can be realized by multiplying each vector field $X_i,\; i\in \left\{+,0,-\right\}$ by a suitable nonvanishing smooth function $f_i,\; i\in \left\{+,0,-\right\}.$


\begin{tw}
The Lie module $\left\{|X_-,X_0,X_+|\right\}$ corresponding to the Euler system (\ref{euler}) can be transformed by an angle-preserving transformation into the unique (up to isomorphism) real Lie algebra $\mathcal{A}$ or $\mathcal{H}.$ Moreover, this Lie algebra is  isomorphic with respect to the Lie algebra $K^-$ or $K^+$ determined by Theorem \ref{infimp}. 
\begin{dow}    
We consider only one instance of  decomposition. The proof for a second decomposition follows by analogy.

Let us define a new basis of vector fields $$\bar{X}_0 := fX_0,\quad \bar{X}_- := gX_-,\quad \bar{X}_+ := hX_+,$$ where $f,g,h \in C^{\infty}$ and $f,g,h \neq 0.$

We compute the Lie bracket $\left[\bar{X}_-,\bar{X}_+\right]$ and express the obtained vector fields in terms of the basis $X_+, X_-, X_0.$ That is,
\begin{equation}
    \begin{aligned}
        \left[\bar{X}_-,\bar{X}_+\right] = \left[gX_-,hX_+\right]
        =gX_-\left(h\right)X_+ - hX_+\left(g\right)X_-+hg\left[X_-,X_+\right] \\= -\frac{h}{g}X_+\left(g\right) \bar{X}_- + \frac{g}{h}X_-\left(h\right)\bar{X}_+. 
    \end{aligned}
\end{equation}
In this way, we obtain the first three structure constants
$$c_1 =0,\quad c_2 = -\frac{h}{g}X_+\left(g\right),\quad c_3 = \frac{g}{h}X_-\left(h\right).$$

\noindent We proceed similarly with the other Lie bracket
\begin{equation}
    \begin{aligned}
        \left[\bar{X}_0,\bar{X}_+\right]=\frac{1}{f}\left(fh-hX_+\left(f\right)\right)\bar{X}_0 + \frac{fh}{g}\frac{1}{4\rho}\bar{X}_- + \frac{1}{h}\left(fX_0\left(h\right)-\frac{fh}{4\rho}\right)\bar{X}_+,
    \end{aligned}
\end{equation}
for which the corresponding structure constants are $$c_4=\frac{1}{f}\left(fh-hX_+\left(f\right)\right),\quad c_5 = \frac{fh}{g}\frac{1}{4\rho},\quad c_6 = \frac{1}{h}\left(fX_0\left(h\right)-\frac{fh}{4\rho}\right).$$
The third Lie bracket
\begin{equation}
    \begin{aligned}
        \left[\bar{X}_0,\bar{X}_-\right]= \frac{1}{f}\left(fg-gX_-\left(f\right)\right)\bar{X}_0 + \frac{1}{g}\left(fX_0\left(g\right)-\frac{fg}{4\rho}\right)\bar{X}_- + \frac{fg}{h}\frac{1}{4\rho}\bar{X}_+,
    \end{aligned}
\end{equation}
gives the structure constants
$$c_7 = \frac{1}{f}\left(fg-gX_-\left(f\right)\right),\quad c_8=\frac{1}{g}\left(fX_0\left(g\right)-\frac{fg}{4\rho}\right),\quad c_9=\frac{fg}{h}\frac{1}{4\rho}.$$

\noindent We obtain a system of nine PDEs determining the structure constants $c_i,\; i\in \left\{1,\dots,9\right\}.$ Now we determine all possible triples $f,g,h \in C^{\infty}$ satisfying this system of equations and all possible values of the structure constants $c_i,\; i\in \left\{1,\dots,9\right\},$
 which we will refer to as  $c_i$-equations. From the equation determining $c_5$ we obtain $$h = \frac{4\rho g c_5}{f}.$$ Applying this formula to the equation for $c_9$ we get $$f = \pm 4\rho c^{\frac{1}{2}}_5 c^{\frac{1}{2}}_9,$$  and $$h = \pm g c^{\frac{1}{2}}_5 c^{-\frac{1}{2}}_9.$$ Now it is easy to verify that $c_6 = c_8.$ 

Applying the above formula for the function $f$ to the equation for $c_4$, we calculate that $c_4=0.$ This further implies that $X_+\left(f\right) = f.$ Proceeding similarly with the $c_7$-equation, we obtain $c_7=0.$

As $c_2 = \mp c^{\frac{1}{2}}_5 c^{-\frac{1}{2}}_9 X_+ \cdot \nabla g$ and $c_3 = X_- \cdot \nabla g$ we obtain
$$c_2\left(\mp c^{-\frac{1}{2}}_5 c^{\frac{1}{2}}_9\right)-c_3 = \left(X_+ - X_-\right)\cdot \nabla g = 2p^{\frac{1}{2}}\rho^{-\frac{1}{2}}\frac{\partial g}{\partial u}.$$ This gives 
$$\frac{\partial g}{\partial u} = p^{-\frac{1}{2}}\rho^{\frac{1}{2}}\frac{1}{2}\left(c_2\left(\mp c^{-\frac{1}{2}}_5c^{\frac{1}{2}}_9\right)-c_3\right),$$ and thus 
$$g = p^{-\frac{1}{2}}\rho^{\frac{1}{2}}\frac{1}{2}\left(c_2\left(\mp c^{-\frac{1}{2}}_5c^{\frac{1}{2}}_9\right)-c_3\right)u+c\left(\rho,p\right),$$ where the function $c\left(\rho,p\right)$ is $u$-independent. To shorten the notation, we write $c=c\left(\rho,p\right).$ 
The above form of the function $g$ implies that the function $h$ is given by $$h =\pm c^{\frac{1}{2}}_5c^{-\frac{1}{2}}_9\left(p^{-\frac{1}{2}}\rho^{\frac{1}{2}}\frac{1}{2}\left(c_2\left(\mp c^{-\frac{1}{2}}_5c^{\frac{1}{2}}_9\right)-c_3\right)u+c\left(\rho,p\right)\right).$$

\noindent Let us observe that as $c_7=0$ we have  $X_-\left(f\right) = f,$ thus $$X_+\left(f\right) = X_-\left(f\right).$$ This equality immediately implies that $$\frac{\partial f}{\partial u}\equiv 0.$$

\noindent Applying the above form of the function $f$ to the $c_8$-equation, we obtain
$$c_8 = \pm 4\rho c^{\frac{1}{2}}_5c^{\frac{1}{2}}_9 \frac{\partial \ln g}{\partial \rho}\mp c^{\frac{1}{2}}_5 c^{\frac{1}{2}}_9,$$ which gives
$$c_8\pm c^{\frac{1}{2}}_5c^{\frac{1}{2}}_9 = \pm 4\rho c^{\frac{1}{2}}_5c^{\frac{1}{2}}_9 \frac{\partial \ln g}{\partial \rho}.$$
In this way, we obtain
$$\frac{\partial \ln g}{\partial \rho} = \frac{1}{4\rho}\frac{c_8 \pm c^{\frac{1}{2}}_5c^{\frac{1}{2}}_9}{\pm c^{\frac{1}{2}}_5c^{\frac{1}{2}}_9},$$ and after rewriting a derivative of the logarithm, and by using the fact that
$$\frac{\partial g}{\partial \rho} = \frac{1}{4}p^{-\frac{1}{2}}\rho^{-\frac{1}{2}}\left(\mp c_2c^{-\frac{1}{2}}_5c^{\frac{1}{2}}_9-c_3\right)u+ \frac{\partial c}{\partial \rho},$$
we obtain  
$$A\left(\frac{1}{8}p^{-\frac{1}{2}}\rho^{-\frac{1}{2}}\left(c_2\left(\mp c^{-\frac{1}{2}}_5c^{\frac{1}{2}}_9\right)-c_3\right)u+\frac{c}{4\rho}\right) = \frac{1}{4}p^{-\frac{1}{2}}\rho^{-\frac{1}{2}}\left(c_2\left(\mp c^{-\frac{1}{2}}_5c^{\frac{1}{2}}_9\right)-c_3\right)u + \frac{\partial c}{\partial \rho},$$
where the constant $A$ is $$A = \frac{c_8\pm c^{\frac{1}{2}}_5c^{\frac{1}{2}}_9}{\pm c^{\frac{1}{2}}_5c^{\frac{1}{2}}_9}.$$
This leads to the equality 
$$\frac{8}{2-A}p^{\frac{1}{2}}\rho^{\frac{1}{2}}\left(A\frac{c}{4\rho}-\frac{\partial c}{\partial \rho}\right) = \left(\mp c_2c^{-\frac{1}{2}}_5c^{\frac{1}{2}}_9-c_3\right)u.$$ The left-hand side of the above equality does not depend on the $u$-variable, therefore, the right-hand side should also be $u$-independent. So we have $\mp c_2c^{-\frac{1}{2}}_5c^{\frac{1}{2}}_9 -c_3 =0.$ This leads to $$\frac{\partial g}{\partial u} =0\quad \text{ and }\quad g = c\left(\rho,p\right),$$ 
and $$\frac{\partial h}{\partial u}=0,\quad \text{ and }\quad h= \pm c^{\frac{1}{2}}_5c^{-\frac{1}{2}}_9 c\left(\rho,p\right).$$

\noindent The $c_8$-equation implies that 
$$\frac{\partial \ln c}{\partial \rho}= B\frac{1}{\rho},$$ where $B$ is a constant given by $$B=\frac{c_8c^{-\frac{1}{2}}_5c^{-\frac{1}{2}}_9\pm 1}{\pm 4}.$$ The above equation can be solved explicitly, so we have $$c = \rho^B e^{c'},$$ where $c'=c'\left(p\right).$ Using the obtained form of the function $\frac{\partial c}{\partial \rho}$ and inserting it into the $c_2$-equation, we get 
$$c_2\left(\mp c^{-\frac{1}{2}}_5 c^{\frac{1}{2}}_9\right)=c \frac{c_8c^{-\frac{1}{2}}_5c^{-\frac{1}{2}}_9\pm1}{\pm 4}+p\frac{\partial c}{\partial p}.$$
Using the above formula for the function $c$ we rectify this equality to get
$$\rho^{-B}c_2\left(\mp c^{-\frac{1}{2}}_5c^{\frac{1}{2}}_9\right)=e^{c'}\left(B+p\frac{\partial c'}{\partial p}\right).$$ Let us observe that the right-hand side of the above is $\rho$-independent, thus $B=0.$ This means that
$$0 = c_8c^{-\frac{1}{2}}_5c^{-\frac{1}{2}}_9 \pm 1,\quad 
c=e^{c'}\; \text{ and }\; 
c_2\left(\mp c^{-\frac{1}{2}}_5c^{\frac{1}{2}}_9\right)=e^{c'}p\frac{\partial c'}{\partial p}.$$ Let us denote $J = c_2\left(\mp c^{-\frac{1}{2}}_5c^{\frac{1}{2}}_9\right).$ Then
$$\frac{1}{p}J = \frac{\partial}{\partial p}\left(e^{c'}\right).$$ Solving this we conclude that $$c = J\ln p + \widetilde{c},$$ where $\widetilde{c}$ is a constant. 

Summarizing the obtained results
$$f = \pm 4 \rho c^{\frac{1}{2}}_5c^{\frac{1}{2}}_9,\quad g=c,\quad h=\pm c^{\frac{1}{2}}_5c^{-\frac{1}{2}}_9 c,\quad c = c\left(p\right) = J\ln p + \widetilde{c},$$ let us note that the structure constants which are not zero are expressed in terms of the unconstrained constants $c_5$ and $c_9:$ 
\begin{equation}
    \begin{aligned}
        c_1=0,\quad c_2 = \mp c^{\frac{1}{2}}_5c^{-\frac{1}{2}}_9J,\quad c_3 = c_2\left(\mp c^{-\frac{1}{2}}c^{\frac{1}{2}}_9\right),\\
        c_4=0,\quad
        c_6 = \mp c^{\frac{1}{2}}_5c^{\frac{1}{2}}_9,\quad
        c_7=0,\quad c_8=\mp c^{\frac{1}{2}}_5 c^{\frac{1}{2}}_9. 
    \end{aligned}
\end{equation}

\noindent Therefore, a rescaled Lie algebra $\left\{\bar{X}_0,\bar{X}_-,\bar{X}_+\right\}$ takes the form 
\begin{equation}
    \begin{aligned}
        &\left[\bar{X}_-,\bar{X}_+\right]=c_2\bar{X}_-\mp c_2c^{-\frac{1}{2}}_5c^{\frac{1}{2}}_9\bar{X}_+,\quad
        \left[\bar{X}_0,\bar{X}_+\right]=c_5\bar{X}_- \mp c^{\frac{1}{2}}_5c^{\frac{1}{2}}_9\bar{X}_+,\\
        &\left[\bar{X}_0,\bar{X}_-\right]=\mp c^{\frac{1}{2}}_5c^{\frac{1}{2}}_9\bar{X}_- + c_9\bar{X}_+.
    \end{aligned}
\end{equation}

Our aim is to determine the structure of this Lie algebra. Denoting $c_5=a,\; c_9=b,\; c_2=c$ and further $\alpha=a,\; \beta = a^{\frac{1}{2}}b^{\frac{1}{2}}$, we obtain
\begin{equation}
    \begin{aligned}
        &\left[\bar{X}_-,\bar{X}_+\right]=\frac{c}{\alpha}\left(\alpha\bar{X}_- \mp \beta \bar{X}_+\right),\quad
        \left[\bar{X}_0,\bar{X}_+\right]=\alpha \bar{X}_- \mp \beta \bar{X}_+,\\ 
        &\left[\bar{X}_0,\bar{X}_-\right]=-\frac{\beta}{\alpha}\left(\pm \alpha \bar{X}_- - \beta \bar{X}_+\right).
    \end{aligned}
\end{equation}

\noindent Let us put $Z_- := \alpha \bar{X}_-$ and $Z_+ := \beta \bar{X}_+.$ The algebra $\left\{\bar{X}_0,Z_-,Z_+\right\}$ takes the form
\begin{equation}\label{ab1}
    \begin{aligned}
        \left[Z_-,Z_+\right]= \beta c\left(Z_- \mp Z_+\right),\quad
        \left[\bar{X}_0,Z_+\right]=\beta \left(Z_- \mp Z_+\right),\quad
        \left[\bar{X}_0,Z_-\right] = - \beta\left(\pm Z_- - Z_+\right).
    \end{aligned}
\end{equation}

\noindent Substituting $Z_0 := \frac{1}{\beta}\bar{X}_0$ and $r := \beta c$ into \eqref{ab1} leads us to the algebra $\left\{Z_0,Z_-,Z_+\right\}:$
\begin{equation}
    \begin{aligned}
        \left[Z_-,Z_+\right]= r\left(Z_-\mp Z_+\right),\quad
        \left[Z_0,Z_+\right]= Z_- \mp Z_+,\quad
        \left[Z_0,Z_-\right] = \mp Z_- + Z_+.
    \end{aligned}
\end{equation}
Different choices of signs provide the following two cases. The `plus' case
\begin{equation}
    \begin{aligned}
        \left[Z_-,Z_+\right]= r\left(Z_- + Z_+\right),\quad
        \left[Z_0,Z_+\right]= Z_- + Z_+,\quad
        \left[Z_0,Z_-\right] =  Z_- + Z_+.
    \end{aligned}
\end{equation}
and the `minus' case
\begin{equation}
    \begin{aligned}
        \left[Z_-,Z_+\right]= r\left(Z_- - Z_+\right),\quad
        \left[Z_0,Z_+\right]= Z_- - Z_+,\quad
        \left[Z_0,Z_-\right] =  -Z_- + Z_+.
    \end{aligned}
\end{equation}
It is easy to see that these two choices give isomorphic Lie algebras. Indeed, a function $\Phi$ maps the `plus' case into the `minus' case and $\Phi$ given by $$\Phi\left(Z_0\right)=-Z_0,\quad \Phi\left(Z_-+Z_+\right)= Z_+-Z_-,\quad \Phi\left(Z_+ - Z_-\right)= Z_- + Z_+,$$ is an isomorphism of these two Lie algebras. Thus, we can restrict our attention to one case, say, to the `plus' case.

Let us consider a new basis $\alpha_0 := Z_0,\; \alpha_1 := Z_+-Z_-,\; \alpha_2 := Z_-+Z_+.$ Then
\begin{equation}
    \begin{aligned}
        \left[\alpha_0,\alpha_1\right] = 0,\quad
        \left[\alpha_0,\alpha_2\right] = 2\alpha_2,\quad
        \left[\alpha_1,\alpha_2\right]= - 2r\alpha_2.
    \end{aligned}
\end{equation}
Now, by a change of basis of the form $\zeta_0:=\frac{1}{2}\alpha_0,\; \zeta_1 := \alpha_1+r\alpha_0,\; \zeta_2 := \alpha_2,$ we obtain 
\begin{equation}
    \begin{aligned}
        \left[\zeta_1,\zeta_2\right] =0,\quad
        \left[\zeta_0,\zeta_2\right]=\zeta_2,\quad
        \left[\zeta_0,\zeta_1\right]=0.
    \end{aligned}
\end{equation}
Rewriting for clarity the vector fields $\zeta_0, \zeta_1, \zeta_2$ in the basis $\bar{X}_0,\bar{X}_-,\bar{X}_+,$ we obtain 
\begin{equation}\label{final1}
\zeta_0 = \frac{1}{2\beta}\bar{X}_0,\quad \zeta_1= \beta \bar{X}_+ - \alpha\bar{X}_-+ \frac{r}{\beta}\bar{X}_0,\quad \zeta_2 = \alpha \bar{X}_- + \beta \bar{X}_+.
\end{equation}
For further applications, it is important to recall that, up to  multiplication by a real constant, we have 
$$\bar{X}_+ \approx \left(\ln p + \widetilde{c}\right)X_+,\quad \bar{X}_- \approx \left(\ln p + \widetilde{c}\right) X_-,\quad \bar{X}_0 \approx \rho X_0.$$
As our consideration covered all possible algebras obtained by rescalings, we conclude that the Lie module $\left\{|X_-,X_0,X_+|\right\}$ is rescaled to the unique (up to isomorphism) real Lie algebra $\left\{Y_0,Y_1,Y_2\right\},$ which satisfies the Jacobi identity. Finally, let us point out that the constants $c_5$ and $c_9$ need to be different from zero due to the assumption that the rescaling functions are nonvanishing. The structure constant $c_2$ can be chosen to be zero. This completes the proof. \qed 
\end{dow}
\end{tw}
\begin{uw}
In the proof, we established the uniqueness of the real Lie algebra by exhausting all possibilities. Further, in Section \ref{infinite-dimensional Lie algebras}, we prove a general uniqueness theorem for related transformations of modules. 
\end{uw}

\subsection{Determining a parametrization  of non-elastic wave superpositions  }\label{7.1}

Let $X_-, X_0, X_+$ be eigenvectors of the one-dimensional Euler system \eqref{euler} satisfying the commutator relation (\ref{eulercom2}).
For the sake of simplicity of computations, we assume that $\kappa=3,$ but the case $\kappa>0$ follows analogously. 

According to Definition \ref{qfam}, the family $\left\{X_-,X_0,X_+\right\}$ is not quasi-rectifiable and the subfamily $\left\{X_-,X_+\right\}$ is quasi-rectifiable. The quasi-rectifiability of the latter family implies that $X_-$ and $X_+$ span a certain two-dimensional manifold $S.$ By the straightening of vector fields theorem (Theorem \ref{straightening}) there exists a certain parametrization $g=g\left(r_1,r_2\right)$ of the surface $S$ such that $$\frac{\partial g}{\partial r_1} = hX_+\quad \text{ and }\quad \frac{\partial g}{\partial r_2} = hX_-,$$ where the function $h$ is a scaling factor and is the same for both vector fields $X_-$ and $X_+,$ namely $$h = \frac{1}{2\sqrt{3}}\rho^{\frac{1}{2}}p^{-\frac{1}{2}}.$$ Our objective is to describe an interaction of three waves in the region $\mathcal{M}$ of superposition  in terms of the flow of the vector field $X_0$ through the surface $S$ spanned by $X_-$ and $X_+.$ We aim to express a superposition of waves in the Euler system \eqref{euler} as an evolution of the surface $S.$ 

As the family $\left\{X_-, X_0, X_+\right\}$ is not quasi-rectifiable, we cannot use the Rescaling Theorem \eqref{algebra-ic_2} to obtain the proper parametrization of the region of wave superposition. Let us recall the fact that the  property of quasi-rectifiability  depends on the choice of basis. Therefore, we now aim at finding an equivalent formulation for the vector fields ${X_+, X_-, X_0}$, which satisfies the assumptions of Theorem 2.3. 

By the rescaling result of the previous section, we know that the quasi-rectifiable basis of the Lie module $\left\{|X_-,X_0,X_+|\right\}$ exists and takes the form of the family \eqref{final1}.  To avoid computational difficulties, instead of the basis $\eqref{final1},$ we consider a slightly different one. 
Let us define $$Z_1:=X_++X_-,\quad Z_2:= X_+-X_-,$$ and consider the family $\left\{Z_1, Z_2, X_0\right\}.$ Note that the vector fields $Z_1, Z_2, X_0$ are linearly independent, i.e., 
\begin{equation}
Z_1 \wedge Z_2 \wedge X_0 \neq 0\quad \text{ at any point in } \mathbb{R}^3. 
\end{equation}
Moreover, we can easily check that 
$\spam\left\{X_-,X_+\right\}=\spam\left\{Z_1,Z_2\right\}.$
These facts mean that $\left\{Z_1,Z_2\right\}$ lies on the same surface $S$ as $\left\{X_-,X_+\right\}.$ In the coordinates $\left\{\rho, p, u\right\}$ introduced in \eqref{euler} the vector fields $Z_1,Z_2,X_0$ take the form
\begin{equation}
\begin{aligned}
Z_1=\left(2\rho,6p,0\right),\quad 
Z_2=\left(0,0,2\sqrt{3}p^{\frac{1}{2}}\rho^{-\frac{1}{2}}\right)=2 \sqrt{3}p^{\frac{1}{2}}\rho^{-\frac{1}{2}} e_3,\quad 
X_0 = \left(1,0,0\right) = e_1. 
\end{aligned}
\end{equation}
Let us compute the Lie brackets of the vector fields $Z_1,Z_2$ and $X_0:$ 
\begin{equation}\label{kom}
\begin{aligned}
\left[Z_1,Z_2\right]=2Z_2,\quad
\left[X_0,Z_1\right]=2X_0,\quad
\left[X_0,Z_2\right]=-\frac{1}{2}\rho^{-1}Z_2.
\end{aligned}
\end{equation}
Note that the relation $\left[X_0,Z_2\right]=-\frac{1}{2}\rho^{-1}Z_2$ implies that $\left\{Z_1,Z_2,X_0\right\}$ is not a finite-dimensional real Lie algebra, but a $C^{\infty}$-Lie module which can be identified with  an infinite-dimensional real Lie algebra. The commutation relations \eqref{kom} imply that the family of vector fields $\left\{Z_1,Z_2,X_0\right\}$ is in quasi-rectifiable form. 

Let us determine the rescaled vector fields $h_1Z_1, h_2Z_2$ and $h_0X_0$ that commute among themselves, i.e.,
\begin{equation}\label{scaled}
\begin{aligned}
\left[h_0X_0,h_1Z_1\right]=0,\quad
\left[h_0X_0, h_2Z_2\right]=0,\quad 
\left[h_1Z_1, h_2Z_2\right]=0.
\end{aligned}
\end{equation}
The dual $1$-forms $\eta_1, \eta_2$ and $\eta_0$ associated with the vector fields $Z_1, Z_2, X_0,$ respectively, are  $$\eta_1\left(\rho,p,u\right)=\frac{1}{6}p^{-1}dp,\quad \eta_2\left(\rho,p,u\right)= \frac{1}{2\sqrt{3}}p^{-\frac{1}{2}}\rho^{\frac{1}{2}}du,\quad \eta_0\left(\rho,p,u\right)=d\rho - \frac{1}{3}\rho p^{-1} dp.$$  The $1$-form $\eta_1$ is already closed, i.e., $d\eta_1 = 0.$
The two $1$-forms $\eta_2$ and $\eta_0$ are not in a closed form and we look for integrating factors for these $1$-forms. We find that $$d\left(p^{\frac{1}{2}}\rho^{-\frac{1}{2}}\eta_2\right) = 0.$$ Thus the integrating factor for $\eta_2$ can be chosen as $j_2:=p^{\frac{1}{2}}\rho^{-\frac{1}{2}}.$ Similarly, we obtain that $d\left(\rho^{-1}\eta_0\right)=0.$ Let us define $j_0:=\rho^{-1}.$ Then  the rescaling functions are
\begin{equation}
\begin{aligned}
h_1:=1,\quad 
h_2:=\frac{1}{j_1}= p^{-\frac{1}{2}}\rho^{\frac{1}{2}},\quad 
h_0:=\rho.
\end{aligned}
\end{equation}
It can easily be checked that such functions $h_0,h_1,h_2$ are proper scaling factors (\eqref{scaled} holds) for the vector fields $Z_1, Z_2, X_0,$ respectively. In what follows, we denote 
\begin{equation}\label{bar1}
\begin{aligned}
\bar{Z}_1 := Z_1,\quad 
\bar{Z}_2 := h_2Z_2=2\sqrt{3}e_3,\quad 
\bar{X}_0:=h_0 X_0 = \rho e_1.
\end{aligned}
\end{equation}
Since all Lie brackets of the vector fields \eqref{bar1} vanish, the wave superposition manifold $v\left(\mathcal{M}\right)$ (corresponding to the non-elastic interaction) can be parametrized by certain functions, say $t_1,t_2,t_3$. Then, the unknown function $v=\left(\rho, p, u\right)$ can be expressed in the form 

\begin{equation} \label{param}
    v\left(\mathcal{M}\right)\colon \hspace{2mm}  v=f\left(t_1,t_2,t_3\right)=\left(f^1\left(t_1,t_2,t_3\right), f^2\left(t_1,t_2,t_3\right),f^3\left(t_1,t_2,t_3\right)\right) \in \mathbb{R}^3
\end{equation}
satisfying the partial differential equations $$\frac{\partial f}{\partial t_1} = Z_1\left(f\right),\quad \frac{\partial f}{\partial t_2} = \bar{Z}_2\left(f\right)=2\sqrt{3}e_3,\quad \frac{\partial f}{\partial t_3} = \bar{X}_0\left(f\right),$$ which leads to nine equations: 
\begin{equation}\label{nine}
\begin{aligned}
&\frac{\partial f^1}{\partial t_1} = 2f^1,\quad  \frac{\partial f^1}{\partial t_2} = 0,\quad \frac{\partial f^1}{\partial t_3} = f^1,\quad
\frac{\partial f^2}{\partial t_1} = 6f^2,\quad \frac{\partial f^2}{\partial t_2} = 0,\quad \frac{\partial f^2}{\partial t_3} = 0,\\
&\frac{\partial f^3}{\partial t_1} = 0,\quad \frac{\partial f^3}{\partial t_2} = 2\sqrt{3},\quad \frac{\partial f^3}{\partial t_3} = 0. 
\end{aligned}
\end{equation}
A general solution of \eqref{nine} has the form
\begin{equation}\label{para}
v=\left(\rho,p,u\right)=f\left(t_1,t_2,t_3\right)=\left(e^{2t_1+t_3},e^{6t_1},2\sqrt{3}t_2\right).
\end{equation}
\begin{uw}
Let us observe that it is not possible to find a nontrivial function $$\alpha = \left(\alpha^1\left(z_1,z_2,z_3\right),\alpha^2\left(z_1,z_2,z_3\right),\alpha^3\left(z_1,z_2,z_3\right)\right),$$ which satisfies the  rectified system of equations 
\begin{equation}
\begin{aligned}
Z_1\left(\alpha\right)=h^{-1}\left(\alpha\right)\left(\frac{\partial \alpha}{\partial z_1}+\frac{\partial \alpha}{\partial z_2}\right),\quad
\bar{Z}_2\left(\alpha\right)=h^{-1}\left(\alpha\right)\left(\frac{\partial \alpha}{\partial z_1}-\frac{\partial \alpha}{\partial z_2}\right),\quad
\frac{\partial \alpha}{\partial z_3}=\bar{X}_0\left(\alpha\right).
\end{aligned}
\end{equation}
The existence of such a function would give the explicit solution of the compressible Euler system. However, it can be shown that the only function that satisfies this system of equations is the zero function.
\end{uw}

If we fix the third variable $t_3$ in \eqref{para}, say $t_3=c$, then $f\left(t_1,t_2,c\right)= \left(e^c e^{2t_1},e^{6t_1},2\sqrt{3}t_2\right)$ defines the family of surfaces $S$ shifted in the $\rho$-direction by the factor $e^c.$ Moreover, if we fix the variable $t_2$ in \eqref{para}, that is $t_2=b,$ we obtain 
$f\left(t_1,b,t_3\right)=\left(e^{2t_1+t_3},e^{6t_1},0\right) + \left(0,0,2\sqrt{3}b\right).$ 
In this case, the vector fields $\left(e^{2t_1+t_3},e^{6t_1},0\right)\; \text{ and }\; \left(0,0,2\sqrt{3}b\right)$ are orthogonal to each other in $\mathbb{R}^3.$

\subsection{Deformations of the quasi-rectifiable surface}
{We study the geometric structures related to the given family of quasi-rectifiable vector fields. We propose a description of the manifold of superposition $v\left(\mathcal{M}\right)$ in terms of a flow of the surface related to the subfamily of vector fields $\left\{X_-,X_+\right\}.$ We prove that, after suitable transformations, deformations of this quasi-rectifiable surface can be expressed as a parallel transport.}

\begin{defi}
Let $\left\{X_1,\dots,X_k\right\}$ be a family of linearly independent vector fields such that the subfamily $\left\{X_1,\dots,X_l\right\},\; l\leqslant k$ is quasi-rectifiable. An $l$-dimensional manifold $M$ such that the tangent bundle of $M$ is spanned by the vector fields $X_1,\dots,X_l,$ that is, $TM = \spam\left\{X_1,\dots,X_l\right\},$ is called a quasi-rectifiable manifold. In the special case of the two-dimensional quasi-rectifiable subfamily $\left\{X_1,X_2\right\}$, a $2$-dimensional manifold $M$ such that $TM = \spam\left\{X_1,X_2\right\}$ is called a quasi-rectifiable surface.   
\end{defi}

From the computational point of view, it is crucial to determine the simplest form of the vector fields that span the same geometric structure. To this end, we propose the following definition.
\begin{defi}
Let $\left\{X_1,\dots,X_l\right\}$ and $\left\{Y_1,\dots,Y_l\right\}$ be two quasi-rectifiable families of vector fields. If the $l$-dimensional manifold $M$ is a quasi-rectifiable manifold of the families $\left\{X_1,\dots,X_l\right\}$ and $\left\{Y_1,\dots,Y_l\right\},$ i.e., $TM=\spam\left\{X_1,\dots,X_l\right\}$ and $TM=\spam\left\{Y_1,\dots,Y_l\right\}$ then we say that the families $\left\{X_1,\dots,X_l\right\}$ and $\left\{Y_1,\dots,Y_l\right\}$ are geometrically related.    
\end{defi}

Let us consider a new parametrization defined by 
\begin{equation}
\begin{aligned}
\beta\left(t_1,t_2,t_3\right):= \ln f\left(t_1,t_2,t_3\right) = \left(2t_1+t_3,6t_1, \ln2\sqrt{3} + \ln t_2\right)\\
= \left(2t_1,6t_1, \ln2\sqrt{3} + \ln t_2\right) + \left(t_3,0,0\right). 
\end{aligned}
\label{parametrisation}\end{equation} 
Let us denote the three-dimensional manifold parametrized by the function $f$ as
\begin{equation}
    S_1 := \left\{v=\left(\rho, p, u \right)=f\left(t_1,t_2,t_3\right)= \left(e^{2t_1+t_3},e^{6t_1},2\sqrt{3}t_2\right)\colon  t_1,t_2,t_3 \in \mathbb{R}_+\right\},
\end{equation}
and its image via the logarithm map as
\begin{equation}
    S_2 := \left\{\beta\left(t_1,t_2,t_3\right)= \left(2t_1+t_3,6t_1,\ln2\sqrt{3}+\ln t_2\right)\colon  t_1,t_2,t_3 \in \mathbb{R}_+\right\}.
\end{equation}

The  vector fields tangent to the manifold $S_2$, corresponding to the differentiation of $\beta$ with respect to $t_i,\; i\in \left\{1,2,3\right\}$ are 
\begin{equation}
\begin{aligned}
\frac{\partial \beta}{\partial t_1} = \left(2,6,0\right),\quad
\frac{\partial \beta}{\partial t_2} = \left(0,0,\frac{1}{t_2}\right),\quad
\frac{\partial \beta}{\partial t_3} = \left(1,0,0\right).
\end{aligned}
\end{equation}
Now we can easily compute the flow $\phi$ of the vector field $\frac{\partial \beta}{\partial t_3}$ by solving the initial value problem
\begin{equation}\label{flow1}
\frac{d \phi}{d t} = \left(1,0,0\right),\quad \phi\left(0\right)=\left(\rho_0,p_0,u_0\right).
\end{equation}
We determine that the solution of \eqref{flow1} takes the form $\phi\left(t\right) = \left(t+\rho_0,p_0,u_0\right).$

Let us verify that the standard Euclidean covariant derivatives $\nabla_{\frac{\partial \beta}{\partial t_3}}\frac{\partial \beta}{\partial t_1}$ and $\nabla_{\frac{\partial \beta}{\partial t_3}}\frac{\partial \beta}{\partial t_2}$ vanish. Indeed,
\begin{equation}\label{parallel}
\begin{aligned}
\nabla_{\frac{\partial \beta}{\partial t_3}}\frac{\partial \beta}{\partial t_1} = \frac{d}{dt}\left(\frac{\partial \beta}{\partial t_1}\left(\phi\left(t\right)\right)\right)= \frac{d}{dt}\left(2,6,0\right)=0,\\
\nabla_{\frac{\partial \beta}{\partial t_3}}\frac{\partial \beta}{\partial t_2} = \frac{d}{dt}\left(\frac{\partial \beta}{\partial t_2}\left(\phi\left(t\right)\right)\right)= \frac{d}{dt}\left(0,0,\frac{1}{p_0}\right)=0.
\end{aligned}
\end{equation}
Let us define the surface spanned by the family $\left\{\dfrac{\partial \beta}{\partial t_1},\dfrac{\partial \beta}{\partial t_2}\right\}$ at a fixed coordinate $t_3=t,$ that is 
\begin{equation}
    \Sigma\left(t\right) := \left\{\left(2t_1 + t, 6 t_1, \ln2\sqrt{3} + \ln t_2\right)\colon  t_1,t_2>0\right\}.
\end{equation}
 The equations \eqref{parallel} imply that the evolution of the surface $\Sigma\left(t\right)$ under the action of the flow of $\dfrac{\partial \beta}{\partial t_3}$ is determined by a parallel transport of the vector fields $\dfrac{\partial \beta}{\partial t_1}$ and $\dfrac{\partial \beta}{\partial t_2}.$ This implies that if $t'>t,$ then $\Sigma\left(t\right) \cap \Sigma\left(t'\right)=\emptyset$ holds. The foliation of the manifold $S_2$ is slicing it into stacks of quasi-rectifiable surfaces (leaves) denoted by $\Sigma(t)$. The manifold takes the form
$$S_2 = \bigcup_{t>0} \Sigma\left(t\right).$$ 
\noindent This means that the foliation of the manifold $v\left(\mathcal{M}\right)$ is slicing it into stacks of quasi-rectifiable surfaces $\Sigma\left(t_3\right)$, often called leaves.

Let us denote 
\begin{equation}\label{fi}
\Phi\left(t\right) := \exp\Sigma\left(t\right) =\left\{ \left(e^{2t_1}e^t,e^{6t_1},2\sqrt{3}t_2\right)\colon  t_1,t_2>0\right\}.
\end{equation}
Note that 
\begin{equation}\label{passage}
\frac{\partial f^j}{\partial t_i} = f^j\frac{\partial \beta^j}{\partial t_i},\; \; i,j \in \left\{1,2,3\right\},
\end{equation}
since $$\frac{\partial f^j}{\partial t_i} = \frac{\partial}{\partial t_i}\left(\exp \beta^j\right) = \exp \beta^j \frac{\partial}{\partial t_i}\beta^j = f^j \frac{\partial \beta^j}{\partial t_i},$$ holds. This means that we can completely reconstruct the evolution of the original surface $\Phi\left(t\right)$ from the parallel transport of the surface $\Sigma\left(t\right).$ Indeed, the following theorem is proven.

\begin{tw}
The transformed manifold $S_2$ related to  superpositions of any two waves associated with the vector fields $\left\{X_-,X_0,X_+\right\}$ can be described as the parallel transport of the quasi-rectifiable surfaces 
$\Sigma\left(t\right),\; t>0,$ and

\begin{equation}S_2 = \bigcup_{t>0}\Sigma\left(t\right),\quad \text{where }\quad\Sigma\left(t\right)\cap \Sigma\left(t'\right)=\emptyset \quad\text{for } t'>t.\end{equation} 

 \noindent Moreover, the manifold $S_1$ can be represented as a deformation of the quasi-rectifiable surface $\Phi\left(t\right),\; t>0.$ The evolution of $\Phi\left(t\right),\; t>0$ can be reconstructed from the evolution of $\Sigma\left(t\right),\; t>0$ by formula \eqref{passage}.
\end{tw}

\subsection{Curvatures of the quasi-rectifiable surface}
{In this subsection, we compute} invariants of a surface spanned by the vector fields $X_-$ and $X_+.$ 

\begin{tw}
Let $t_3>0$ be fixed. Then for the quasi-rectifiable surface $\Phi\left(t_3\right):$ 
\begin{itemize}
    \item[$i)$] the Riemannian metric takes the form \begin{equation}
    g\left(t_1,t_2\right) = \begin{pmatrix}
       4e^{2t_3}e^{4t_1}+ 36e^{12t_1}& 0\\
       0& 12
    \end{pmatrix}, 
    \label{2nd}
\end{equation}
\item[$ii)$] the second fundamental form is \begin{equation}
    II = \begin{pmatrix}
        L & M\\
        M & N
    \end{pmatrix}=
    \begin{pmatrix}
        L&0\\
        0&0
    \end{pmatrix},
\end{equation}
\item [$iii)$] the Gaussian curvature is $K=0,$
\item[$iv)$] the mean curvature is $H=\frac{L}{2}.$
\end{itemize}
\begin{dow}
{We begin by computing the Riemannian metric} $g$ of the quasi-rectifiable surface $\Phi\left(t_3\right)$ (see \eqref{fi}) spanned by the family $\left\{X_-,X_+\right\}.$
{Applying the parametrization $f$ given by formula \eqref{para}, we compute:} 
\begin{equation}
    \begin{aligned}
        \frac{\partial f}{\partial t_1} \cdot \frac{\partial f}{\partial t_1} = 4e^{2t_3}e^{4t_1}+ 36e^{12t_1},\quad
        \frac{\partial f}{\partial t_2}\cdot \frac{\partial f}{\partial t_2}= 12,\quad
        \frac{\partial f}{\partial t_1}\cdot\frac{\partial f}{\partial t_2} = 0.
    \end{aligned}
\end{equation}
From this we derive the form of the Riemannian metric $g$  
\begin{equation}
    g\left(t_1,t_2\right) = \begin{pmatrix}
       4e^{2t_3}e^{4t_1}+ 36e^{12t_1}& 0\\
       0& 12
    \end{pmatrix}.
\end{equation}
Further computations show that the vector field $$n = \frac{\frac{\partial f}{\partial t_1}\times \frac{\partial f}{\partial t_2}}{\big|\frac{\partial f}{\partial t_1}\times \frac{\partial f}{\partial t_2}\big|},$$ normal to $\Phi\left(t_3\right),$ is given by
\begin{equation}
    n = \frac{4\sqrt{3}}{\sqrt{48}}\left(\frac{3e^{4t_1}}{\left(9e^{8t_1}+e^{2t_3}\right)^{\frac{1}{2}}},\frac{e^{t_3}}{\left(9e^{8t_1}+e^{2t_3}\right)^{\frac{1}{2}}},0\right)
\end{equation}
and the second derivatives of the parametrization $f$ are 
\begin{equation}
    \begin{aligned}
        \frac{\partial^2 f}{\partial t_1^2} = \left(4e^{t_3}e^{2t_1},36e^{6t_1},0\right),\quad
        \frac{\partial^2 f}{\partial t_2^2} = \left(0,0,0\right),\quad
        \frac{\partial^2 f}{\partial t_1 \partial t_2}=\left(0,0,0\right).
    \end{aligned}
\end{equation}
Coefficients of the second fundamental form \eqref{2nd} of $\Phi\left(t_3\right)$ can be easily determined to be
\begin{equation}
    \begin{aligned}
        L = \frac{\partial^2 f}{\partial t_1^2} \cdot n = \frac{48 e^{t_3}e^{6t_1}}{\left(9e^{8t_1}+e^{2t_3}\right)^{\frac{1}{2}}},\\
        M = N =0.
    \end{aligned}
\end{equation}

Hence, the eigenvalues of the second form are $k_1 = L$ and $k_2 =0.$
Finally, we obtain the Gaussian curvature $K$ and the mean curvature $H$ for $\Phi\left(t_3\right)$: $$K=k_1k_2 = 0,\qquad H=\frac{1}{2}\left(k_1+k_2\right)=\frac{L}{2}.$$  \qed
\end{dow}
\end{tw}

\subsection{Reduced form of the Euler system}
In what follows we derive a reduced form of the Euler system \eqref{euler} using the parametrization \eqref{para} corresponding to the case of non-elastic wave superpositions. We assume below that $\kappa=3$ (system \eqref{euler} admits additional symmetry \cite{Roz}) but the computations can be similarly performed for any values of $\kappa$.

Let us recall that the parametrization \eqref{para} takes the form
\begin{equation}
f\left(t_1,t_2,t_3\right)=\left(f^1\left(t_1,t_2,t_3,\right),f^2\left(t_1,t_2,t_3,\right),f^3\left(t_1,t_2,t_3,\right)\right)=\left(e^{2t_1+t_3},e^{6t_1},2\sqrt{3}t_2\right).
\label{parametrization}
\end{equation}
{The partial derivatives of the function $f$} with respect to the variables $t_i,\; i\in \left\{1,2,3\right\}$ are
\begin{equation}
    \begin{aligned}
        &\frac{\partial f}{\partial t_1} = \left(2e^{2t_1+t_3},6e^{6t_1},0\right)=h_1\left(f\right)Z_1\left(f\right)=h_1\left(f\right)X_+\left(f\right)+h_1\left(f\right)X_-\left(f\right)=X_+\left(f\right)+X_-\left(f\right),\\
        &\frac{\partial f}{\partial t_2} = \left(0,0,2\sqrt{3}\right)=\bar{w}_2\left(f\right)=h_2\left(f\right)Z_2\left(f\right)=h_2\left(f\right)X_+\left(f\right)-h_2\left(f\right)X_-\left(f\right),\\
        &\frac{\partial f}{\partial t_3} = \left(e^{2t_1+t_3},0,0\right)=\bar{X}_0\left(f\right)=h_0\left(f\right)X_0\left(f\right),
    \end{aligned}
\end{equation}
where $X_-,X_0,X_+$ are the eigenvectors of the Euler system \eqref{euler}{, where $Z_1:=X_++X_-$ and $Z_2:=X_+-X_-$. The rescaling functions are:} $h_1\equiv 1,\; h_2=p^{-\frac{1}{2}}{\rho^{\frac{1}{2}},}\; h_0=\rho$ {for the vector fields $Z_1$, $Z_2$, and $X_0$,} respectively (see Section \ref{7.1} for details). 

\noindent We substitute the function $f$ in the left-hand side (LHS) of the Euler system \eqref{euler} and obtain 
\begin{equation}
    \begin{aligned}
    \text{LHS}={\frac{\partial}{\partial t}}\begin{pmatrix}
        f^1\\
        f^2\\
        f^3
    \end{pmatrix}= 
    \begin{pmatrix}
        \frac{\partial f}{\partial t_1}& \frac{\partial f}{\partial t_2}&\frac{\partial f}{\partial t_3} 
    \end{pmatrix}
    \begin{pmatrix}
        \frac{\partial t_1}{\partial t}\\
        \frac{\partial t_2}{\partial t}\\
        \frac{\partial t_3}{\partial t}
    \end{pmatrix}
    = \begin{pmatrix}
        X_+^1+X_-^1& h_2X_+^1-h_2X_-^1& h_0X_0^1\\
        X_+^2+X_-^2& h_2X_+^2-h_2X_-^2& h_0X_0^2\\
        X_+^3+X_-^3& h_2X_+^3-h_2X_-^3& h_0X_0^3
    \end{pmatrix}
    \begin{pmatrix}
        \frac{\partial t_1}{\partial t}\\
        \frac{\partial t_2}{\partial t}\\
        \frac{\partial t_3}{\partial t}
    \end{pmatrix}\\
    = X_+ \left(\frac{\partial t_1}{\partial t}+ h_2\frac{\partial t_2}{\partial t}\right) + X_-\left(\frac{\partial t_1}{\partial t}-h_2 \frac{\partial t_2}{\partial t}\right) + X_0h_0\frac{t_3}{\partial t}.
\end{aligned}
\end{equation}
In this way we obtain the expression in the basis formed by the vector fields $X_-, X_0, X_+.$
By virtue of the parametric representation of $f$, the right-hand side (RHS) of \eqref{euler} takes the form
\begin{equation}
    \begin{aligned}
    &\text{RHS}=\begin{pmatrix}
        f^3&0&f^1\\
        0&f^3&3f^2\\
        0&\frac{1}{f^1}&f^3
    \end{pmatrix}
    {\frac{\partial}{\partial x}}
    \begin{pmatrix}
        f^1\\
        f^2\\
        f^3
    \end{pmatrix}=
    \begin{pmatrix}
        f^3&0&f^1\\
        0&f^3&3f^2\\
        0&\frac{1}{f^1}&f^3
    \end{pmatrix}
    \begin{pmatrix}
        \nabla f^1 \cdot \frac{\partial \left(t_1,t_2,t_3\right)}{\partial x}\\
        \nabla f^2 \cdot \frac{\partial \left(t_1,t_2,t_3\right)}{\partial x}\\
        \nabla f^3 \cdot \frac{\partial \left(t_1,t_2,t_3\right)}{\partial x}
    \end{pmatrix} \\&= A B 
    \begin{pmatrix}
        \frac{\partial t_1}{\partial x}\\
        \frac{\partial t_2}{\partial x}\\
        \frac{\partial t_3}{\partial x}
    \end{pmatrix}=
    A\left(X_+ \left(\frac{\partial t_1}{\partial x}+ h_2\frac{t_2}{\partial x}\right)+ X_-\left(\frac{\partial t_1}{\partial x}-h_2 \frac{t_2}{\partial x}\right)+ X_0h_0 \frac{\partial t_3}{\partial x}\right),
\end{aligned}
\end{equation}
where we have used the notation
\begin{equation}
    A:=\begin{pmatrix}
        f^3&0&f^1\\
        0&f^3&3f^2\\
        0&\frac{1}{f^1}&f^3
    \end{pmatrix},\quad 
    B := 
    \begin{pmatrix}
        X_+^1+X_-^1& h_2X_+^1-h_2X_-^1& h_0X_0^1\\
        X_+^2+X_-^2& h_2X_+^2-h_2X_-^2& h_0X_0^2\\
        X_+^3+X_-^3& h_2X_+^3-h_2X_-^3& h_0X_0^3
    \end{pmatrix}.
\end{equation}
From the fact that $X_s,\; s \in \left\{-,0,+\right\}$ are linearly independent and $X_s, \lambda_s, \; s\in \left\{-,0,+\right\}$ satisfy the eigenvalue problem for the Euler system \eqref{euler}, we infer that $X_+$ satisfies the equation
\begin{equation}\label{blabla1}
    X_+ \frac{\partial t_1}{\partial t} + X_+h_2\frac{\partial t_2}{\partial t} = \lambda_+ X_+ \frac{\partial t_1}{\partial x} + \lambda_+ X_+ h_2\frac{\partial t_2}{\partial x}.
\end{equation}
In an analogous way, we derive equations for $X_-$ and $X_0.$
By eliminating $X_s,\; s\in \left\{-,0,+\right\}$ from these equations, we obtain the system of partial differential equations
\begin{equation}\label{blabla4}
    \begin{aligned}
        &\frac{\partial t_1}{\partial t} + h_2 \frac{\partial t_2}{\partial t} = \lambda_+ \frac{\partial t_1}{\partial x} + \lambda_+ h_2 \frac{t_2}{\partial x},\quad
        \frac{\partial t_1}{\partial t} - h_2\frac{\partial t_2}{\partial t} = \lambda_- \frac{\partial t_1}{\partial x} - \lambda_- h_2 \frac{\partial t_2}{\partial x},\\
        &h_0 \frac{\partial t_3}{\partial t} = \lambda_0 \frac{\partial t_3}{\partial x}.
    \end{aligned}
\end{equation}
System \eqref{blabla4} can  be writen in the matrix form
\begin{equation}\label{lewy}
    \text{LHS} = 
    \begin{pmatrix}
        1&h_2&0\\
        1&-h_2&0\\
        0&0&h_0
    \end{pmatrix}
    \begin{pmatrix}
        \frac{\partial t_1}{\partial t}\\
        \frac{\partial t_2}{\partial t}\\
        \frac{\partial t_3}{\partial t}
    \end{pmatrix},
\end{equation}
and 
\begin{equation}\label{prawy}
    \text{RHS}=
    \begin{pmatrix}
        \lambda_+&0&0\\
        0&\lambda_-&0\\
        0&0&\lambda_0
    \end{pmatrix}
    \begin{pmatrix}
        \frac{\partial t_1}{\partial x} + h_2 \frac{\partial t_2}{\partial x}\\
        \frac{\partial t_1}{\partial x} - h_2 \frac{\partial t_2}{\partial x}\\
        \frac{\partial t_3}{\partial x}
    \end{pmatrix}=
    \begin{pmatrix}
        \lambda_+&0&0\\
        0&\lambda_-&0\\
        0&0&\lambda_0
    \end{pmatrix}
    \begin{pmatrix}
        1&h_2&0\\
        1&-h_2&0\\
        0&0&h_0
    \end{pmatrix}
    \begin{pmatrix}
        \frac{\partial t_1}{\partial x}\\
        \frac{\partial t_2}{\partial x}\\
        \frac{\partial t_3}{\partial x}
    \end{pmatrix}.
\end{equation}
Comparing \eqref{lewy} with \eqref{prawy} we get the differential equation
\begin{equation}
    \begin{pmatrix}
        \frac{\partial t_1}{\partial t}\\
        \frac{\partial t_2}{\partial t}\\
        \frac{\partial t_3}{\partial t}
    \end{pmatrix}=
    \begin{pmatrix}
        1&h_2&0\\
        1&-h_2&0\\
        0&0&h_0
    \end{pmatrix}^{-1}
    \begin{pmatrix}
        \lambda_+&0&0\\
        0&\lambda_-&0\\
        0&0&\lambda_0
    \end{pmatrix}
    \begin{pmatrix}
        1&h_2&0\\
        1&-h_2&0\\
        0&0&h_0
    \end{pmatrix}
    \begin{pmatrix}
        \frac{\partial t_1}{\partial x}\\
        \frac{\partial t_2}{\partial x}\\
        \frac{\partial t_3}{\partial x}
    \end{pmatrix}.
\end{equation}
After computing the inverse matrix,
we obtain the  form of the above equation with the explicitly given transition matrices \\

\begin{equation}
\frac{\partial}{\partial t}
    \begin{pmatrix}
        t_1\\
         t_2\\
         t_3
    \end{pmatrix}=
    \begin{pmatrix}
        \frac{1}{2}&\frac{1}{2}&0\\
        \frac{1}{2h_2}&-\frac{1}{2h_2}&0\\
        0&0&\frac{1}{h_0}
    \end{pmatrix}
    \begin{pmatrix}
        v_+&0&0\\
        0&v_-&0\\
        0&0&v_0
    \end{pmatrix}
    \begin{pmatrix}
        1&h_2&0\\
        1&-h_2&0\\
        0&0&h_0
    \end{pmatrix}  \frac{\partial }{\partial x}
    \begin{pmatrix}
        t_1\\
       t_2\\
        t_3
    \end{pmatrix}.
\end{equation}
where $h_2\left(f\right)=\left(\frac{\rho}{p}\right)^\frac{1}{2}$ and $h_0\left(f\right)=\rho$ are rescaling functions for the vector fields $X_+-X_-$ and $X_0,$ respectively.

Let us define 

\begin{equation}
    \omega_1=X_++X_-, \quad \omega_2= X_+-X_-, \quad \omega_0=X_0.
\end{equation}

The parametrization \eqref{parametrization} is obtained by integrating the systems of PDEs (see figure 1)

\begin{equation}
    \frac{\partial f}{\partial t_i}=\spam \{\omega_1, \omega_2 \}, \quad i=1,3, \quad \frac{\partial f}{\partial t_2}= \spam \{\omega_0 \}
\end{equation}

\begin{figure}
\centering
\includegraphics[width=0.7\linewidth]{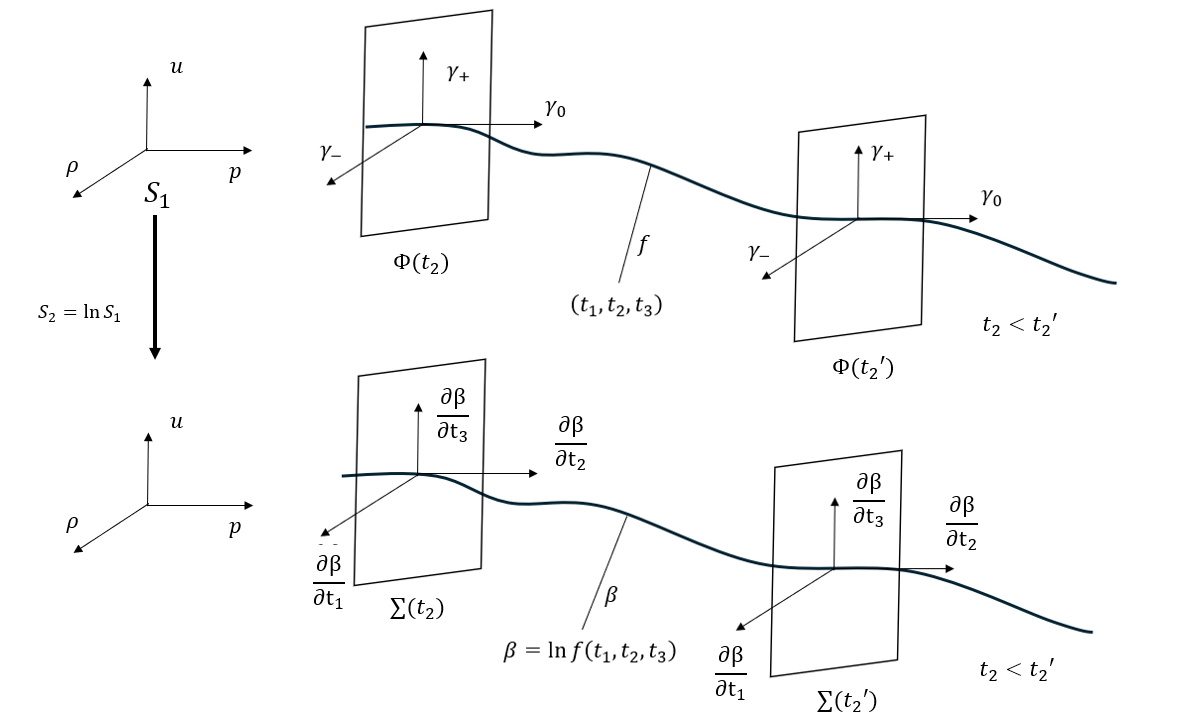}
\caption{The parallel transport of the quasi-rectifiable surfaces $\phi(t_2)$ and $\Sigma(t_2)$ along the trajectory of the vector fields $\gamma_0$ or $\dfrac{\partial\beta}{\partial t_2}$, respectively.}
\label{fig:fig1}
\end{figure}

We summarize the results of this section in the following theorem.
\begin{tw}
If $\left\{|X_-, X_0, X_+|\right\}$ is a Lie module corresponding to the Euler system \eqref{euler}, then the parametrized manifold of a non-elastic wave superposition corresponding to the vector field $X_-, X_0, X_+$ has the form 

\begin{equation*}
v\left(M\right)\colon \quad v=\left(\rho,p,u\right)=f\left(t_1,t_2,t_3\right)=\left(e^{2t_1+t_3},e^{6t_1},2\sqrt{3}t_2\right),\quad \text{for }\kappa=3.
\end{equation*}

The reduced form of the Euler system \eqref{euler} is determined by the hyperbolic system of PDEs of three equations in the three dependent variables $(t_1,t_2,t_3)\in \mathbb{R}^3$ and the two independent variables $(t,x)\in \mathbb{R}^2$

\begin{equation}\label{reduce}
\frac{\partial}{\partial t}\begin{pmatrix}
        t_1\\
        t_2\\
        t_3
    \end{pmatrix}
    -\begin{pmatrix}
        2\sqrt{3} t_2& \sqrt{3}&0\\
        \sqrt{3}e^{4t_1-t_3}&2\sqrt{3}t_2&0\\
        0 &0&2\sqrt{3} t_2
    \end{pmatrix}
    \frac{\partial}{\partial x}
    \begin{pmatrix}
        t_1\\
       	t_2\\
        t_3
    \end{pmatrix}=\begin{pmatrix}
        0\\ 0\\ 0
    \end{pmatrix}.
\end{equation}

Note that the matrix in \eqref{reduce} has the Jordan form and is not diagonalizable unlike the case of the elastic wave superposition \eqref{RI}.

System (\ref{reduce}) constitutes a reduced form of the Euler system (\ref{euler}) corresponding to the case of non-elastic wave superpositions and it is an analog of the reduced system (\ref{RI}) for the elastic case. Note that the matrix in (\ref{reduce}) has the Jordan form and is not diagonalizable unlike the case of (\ref{RI}).  A particular class of explicit solutions of the system \eqref{reduce} is  derived in Appendix B.

\end{tw}

\begin{uw}
Solutions of (\ref{reduce}) describe the process of superposition of an acoustic wave $S_+$ or $S_-$ and an entropic wave E, which results in the production of a new sound wave in the region of interaction. This phenomenon was observed experimentally in wave-particle interactions in plasma physics \cite{Koch, Skoric, Schott}.
\end{uw}

\section{General $C^{\infty}$-Lie modules associated with Riemann waves}

The rescaling result established in the previous section enables a simplification of the structures under consideration, namely those formed by vector fields associated with Riemann waves of the Euler system \eqref{euler}. Specifically, the rescaling transformation allows us to replace the study of $C^{\infty}$-Lie modules of vector fields with the analysis of finite-dimensional real Lie algebras.

This part of the paper is devoted to the generalization of the aforementioned results to an arbitrary Lie module of vector fields, with particular emphasis on collections of vector fields arising from a broad class of hydrodynamic-type systems \eqref{syst_PDEs_2}. In addition to this extension of the previous findings, the present section focuses on an analogous simplification of the underlying geometric structures. In particular, we investigate whether the notions of integrability of a distribution and of tangent manifolds can be reformulated within the framework of Lie groups corresponding to the Lie algebras obtained via the rescaling transformation. By introducing an appropriate affine connection on such Lie groups, we restate the parallel transport result in this more abstract setting. In what follows, our attention is focused on three-dimensional Lie modules of vector fields defined on a three-dimensional manifold.

\subsection{Rescaling transform}
A generic family $\left\{X_1,X_2,X_3\right\}$ of vector fields on a given manifold equipped with the standard Lie bracket spans a Lie module over the ring of smooth functions. By applying suitable conformal (angle-preserving) transformations of a Lie module $\{|X_1,X_2,X_3|\}$ we associate it with  a class of real Lie algebras.

As indicated in the following example, there exist $C^{\infty}$-Lie modules that cannot be rescaled to real Lie algebras.

\begin{exam}
    Let $\{|X_1,X_2,X_3|\}$ be a Lie module over $C^{\infty}.$ Assume that the standard Lie bracket is given by the  relations
    \begin{equation}
        \begin{aligned}
            \left[X_1,X_2\right] = xX_1 + X_3,\quad \left[X_1,X_3\right]=0,\quad \left[X_2,X_3\right]= X_1.
        \end{aligned}
    \end{equation}
    The Jacobi identity implies that $X_3\left(x\right)=0.$
    A simple computation shows that there is no angle-preserving transformation rescaling this module to a real Lie algebra, i.e., there does not exist a triple of nonvanishing smooth functions $f_1,f_2,f_3$ such that $\left\{f_1X_1,f_2X_2,f_3X_3\right\}$ is a real Lie algebra.
\end{exam}

\begin{defi}\label{resc}
A family of nonvanishing real smooth functions $\left\{f_1,f_2,f_3\right\},$ which are associated with  a given $C^{\infty}$-Lie module $\left\{|X_1,X_2,X_3|\right\}$ isomorphic to a real Lie algebra of the form $\left\{|f_1X_1,f_2X_2,f_3X_3|\right\}$ is called a rescaling transformation or simply a rescaling.
\end{defi}

\begin{defi}
    Let $\mathfrak{M}\left(N\right)$ denote the class of three-dimensional $C^{\infty}$-Lie modules of vector fields over the fixed manifold $N.$ By $\mathfrak{M}_{\rm r}\left(N\right)\subset \mathfrak{M}\left(N\right)$ we name a subclass of all $C^{\infty}$-Lie modules for which there exists a rescaling transformation, as in Definition \ref{resc}, to a real Lie algebra.
\end{defi}

As it is shown in  Section \ref{euler-4.2}, the class $\mathfrak{M}_{\rm r}\left(N\right)$ is non-empty -- it contains the module of vector fields associated with the Euler system \eqref{euler}. 

The following theorem specifies conditions characterizing membership in the class $\mathfrak{M}_{\rm r}\left(N\right).$

\begin{tw}
Let $\left\{|X_1,X_2,X_3|\right\} \in \mathfrak{M}\left(N\right)$ be a fixed Lie module with the following bracket relations
\begin{equation}\label{algog}
\begin{aligned}
    &\left[X_1,X_2\right] = aX_1 + bX_2 + cX_3,\quad
    \left[X_1,X_3\right] = dX_1 + eX_2 + fX_3,\\
    &\left[X_2,X_3\right] = gX_1 + hX_2 + iX_3,   
\end{aligned}
\end{equation}
where $a,b,c,d,e,f,g,h,i \in C^{\infty}$ and $c\neq 0.$ 
Then the following statements are equivalent
\begin{itemize}
    \item[$i)$] $\left\{|X_1,X_2,X_3|\right\} \in \mathfrak{M}_{\rm r}\left(N\right),$
    \item[$ii)$] There exist real constants $c_i,\; i\in \{1,\dots,9\}$ such that the system of equations
    \begin{equation}\label{uklad1}
    \begin{aligned}
        &X_1\left(\psi_1\right)=-X_1\left(\ln c\right) + \exp\left(-\psi_1\right)\left(c_6-c_2\right) + b,\qquad X_1\left(\psi_2\right) = c_2 \exp\left(-\psi_1\right) - b,\\ 
        &X_1\left(\psi_3\right) =c_6\exp\left(-\psi_1\right),\qquad
        X_2\left(\psi_1\right)=-c_1\exp\left(-\psi_2\right)+a,\\ 
        &X_2\left(\psi_2\right)=\exp\left(-\psi_2\right)\left(c_1+c_9\right)-X_2\left(\ln c\right)-a,\qquad X_2\left(\psi_3\right)=c_9\exp\left(-\psi_2\right),\\
        &X_3\left(\psi_1\right)=-c_4\exp\left(-\psi_3\right),\qquad X_3\left(\psi_2\right)=-c_8\exp\left(-\psi_3\right),\\ 
        &X_3\left(\psi_3\right)=-\frac{1}{c_3}\left(c_1c_6+c_2c_9\right)\exp\left(-\psi_3\right).
    \end{aligned}
\end{equation}
has a solution $\left\{\psi_1,\psi_2,\psi_3\right\},$ where $\psi_i \in C^{\infty}\left(N\right),\; i\in \left\{1,2,3\right\}.$
\end{itemize}
\begin{dow}
We look for nonvanishing functions $f,g,h\in C^{\infty}$ such that the rescaled vector fields $Y_1 = fX_1,\; Y_2=gX_2,\; Y_3 = hX_3$ form a real Lie algebra. In its general form, the Lie bracket of $\left\{Y_1,Y_2,Y_3\right\}$ can be given as
\begin{equation}\label{stale}
\begin{aligned}
    &\left[Y_1,Y_2\right] = c_1Y_1 + c_2Y_2 + c_3Y_3,\quad
    \left[Y_1,Y_3\right] = c_4Y_1 + c_5Y_2 + c_6Y_3,\\
    &\left[Y_2,Y_3\right] = c_7Y_1 + c_8Y_2 + c_9Y_3,
\end{aligned}
\end{equation}
where $c_i,\; i\in \left\{1,\dots,9\right\}$ are  real structure constants of the Lie algebra.

Applying relations \eqref{algog} to \eqref{stale}, we obtain the system of equations determining the structure constants $c_i,\; i\in \left\{1,\dots,9\right\}:$
\begin{equation}\label{cc}
    \begin{aligned}
        &c_1 = \phi_2\left(a-X_2\left(\ln \phi_1\right)\right),\quad c_2 = \phi_1\left(b+X_1\left(\ln \phi_2\right)\right),\quad c_3=\frac{\phi_1 \phi_2}{\phi_3}c,\\
        &c_4=\phi_3\left(d-X_3\left(\ln \phi_1\right)\right),\quad c_5=\frac{\phi_1\phi_3}{\phi_2}e,\quad c_6=\phi_1\left(f+X_1\left(\ln \phi_3\right)\right),\\
        &c_7=\frac{\phi_2\phi_3}{\phi_1}g,\quad c_8=\phi_3\left(h-X_3\left(\ln \phi_2\right)\right),\quad c_9=\phi_2\left(i+X_2\left(\ln \phi_3\right)\right).
    \end{aligned}
\end{equation}
In the general case, the existence of the rescaling transformation is equivalent to the existence of the smooth functions $\phi_1,\phi_2,\phi_3$ satisfying the system \eqref{cc}. Indeed, by the Jacobi identity for Lie modules, after applying the rescaling transformation, we obtain the set of vector fields, closed over $\mathbb{R}$ with respect to the Lie bracket and satisfying the Jacobi identity, that span a finite-dimensional real Lie subalgebra $\left\{Y_1,Y_2,Y_3\right\}$ of the infinite-dimensional real Lie algebra $\left\{|X_1,X_2,X_3|\right\}.$

Let us denote $\psi_i = \ln \phi_i,\; i\in \{1,2,3\}.$ In the case $c\neq 0,$ we alternatively can rewrite system \eqref{cc} into the form \eqref{uklad1}.
The existence of solutions $\psi_i,\; i\in \left\{1,2,3\right\}$ of this system is equivalent to the existence of a rescaling of the module $\left\{|X_1,X_2,X_3|\right\}.$ The solvability of similar kinds of systems and other existential results are given in \cite{Dar, Mau}, chapter II 23, p.161-240; see also, \cite{Gru2}. \qed
\end{dow}
\end{tw}

\begin{col}
For particular cases of $\left\{|X_1,X_2,X_3|\right\}$ we may provide more direct criteria. Assume that $c,e,g\neq 0.$ From the relations \eqref{stale}, we obtain  
\begin{equation}
    c_3 = \frac{\phi_1\phi_2}{\phi_3}c,\quad c_5=\frac{\phi_1\phi_3}{\phi_2}e,\quad c_7=\frac{\phi_2\phi_3}{\phi_1}g.
\end{equation}
This determines the rescaling functions:
\begin{equation}
    \phi_1=\epsilon_1\left(\frac{c_3c_5}{ce}\right)^{\frac{1}{2}},\quad \phi_2=\epsilon_2\left(\frac{c_3c_7}{cg}\right)^{\frac{1}{2}},\quad \phi_3= \epsilon_1 \epsilon_2 \left(\frac{c_5c_7}{eg}\right)^{\frac{1}{2}},
\end{equation}
where $\epsilon_1, \epsilon_2$ are independent sign functions.
Applying the functions $\phi_1, \phi_2, \phi_3$ to system \eqref{stale}, if this system is not in contradiction, we obtain the differential equations for the structure constants of the real Lie algebra $\left\{Y_1,Y_2,Y_3\right\}.$ 

Similarly, in the case $c,e\neq 0,$ we can determine one rescaling function, for instance, $\phi_1=\epsilon_1\left(\dfrac{c_3 c_5}{ce}\right)^{\frac{1}{2}}$, and then study the solvability of the  system \eqref{cc}. \qed
\end{col}

\begin{col}
If  $c_3=\phi_1 \phi_2 /\phi_3=1$ then using \eqref{algog} we obtain
\begin{equation}
a=X_2 \ln\left(\phi_1\right),\quad b=X_1 \ln\left(\phi_2\right),\quad  d=X_3 \ln\left(X_1\right),\quad  f=-X_1 \ln \left(\phi_1 \phi_2\right).
\end{equation}
This implies that
\begin{equation} 
\begin{aligned}
&c_1=0,\quad  c_2=0,\quad c_3=1,\quad c_4=0,\quad c_5=e,\quad c_6=0,\quad c_7=0,\\ 
&c_8=\phi_1 \phi_2 X_3 \ln \left(\phi_2\right),\quad
c_9=\phi_2 X_2 \ln \left(\phi_1 \phi_2\right).
\end{aligned}
\end{equation}\qed
\end{col}

A powerful technique for establishing the existence of, and an estimate for the degrees of freedom of, solutions of a system of exterior forms in terms of vector fields was established by E. Cartan \cite{Car}. This technique can be used in our approach to determine whether or not a $C^{\infty}$-Lie module of vector fields  can be related to a quasi-rectifiable Lie algebra. This approach can facilitate the analysis of wave superpositions admitted by quasi-rectifiable vector fields.

An example discussed in Appendix \ref{ap6.1} indicates that there exist Lie modules that can be transformed in an angle-preserving way only to a non-quasi-rectifiable real Lie algebra. 

\begin{uw}
It is natural to ask if, for some given Lie module, there exists an alternative angle-preserving transformation rescaling it to a non-isomorphic Lie algebra. The answer is negative and follows from the uniqueness of the rescaling theorem presented in the next section. 

\end{uw}

\subsection{Equivalence classes of rescalings}
We characterize the equivalence classes of the $C^{\infty}$-Lie modules determined by the rescaling transform. 

\begin{lem}
    Let $\left\{|X_1,X_2,X_3|\right\} \in \mathfrak{M}\left(N\right)$ such that $\left[X_i,X_j\right] \in \spam \left\{X_i,X_j\right\}$ for some $\left\{i,j\right\}\subset \left\{1,2,3\right\}.$ Then there exist nonvanishing smooth functions $f,g,h \in C^{\infty}\left(N\right)$ such that $\left\{\bar{X}_1,\bar{X}_2,\bar{X}_3\right\},$ where $\bar{X}_1 = fX_1,\; \bar{X}_2 = gX_2,\; \bar{X}_3 = hX_3$ are such that $$\left[\bar{X}_i, \bar{X}_j\right]=0$$ for any pair $\left\{i,j\right\} \subset \left\{1,2,3\right\}$ satisfying $\left[X_i,X_j\right] \in \spam \left\{X_i,X_j\right\}.$
    \begin{dow}
        A proof follows by a direct application of the Rescaling Theorem \ref{modified}.
        \qed
    \end{dow}
\end{lem}

In the next proposition we show a link between the notion of quasi-rectifiability for Lie modules of vector fields and the quasi-rectifiability of real Lie algebras. 

\begin{tw}
   Assume that $\left\{|X_1,X_2,X_3|\right\} \in \mathfrak{M}_{\rm r}\left(N\right)$ and $\left\{Z_1,Z_2,Z_3\right\}$ is a Lie algebra such that $Z_i = f_iX_i,\; i\in \left\{1,2,3\right\}$ for a rescaling transform $\left\{f_1,f_2,f_3\right\}.$ Then the module $\left\{|X_1,X_2,X_3|\right\}$ is quasi-rectifiable if and only if the Lie algebra $\left\{Z_1,Z_2,Z_3\right\}$ is quasi-rectifiable. 
   \begin{dow}
        As $f_i,\; i \in \left\{1,2,3\right\}$ are nonvanishing we can put $g_i=\frac{1}{f_i},\; i\in \left\{1,2,3\right\}.$
        For any pair of the vector fields $X_i, X_j,\; i,j\in \left\{1,2,3\right\}$ we have 
        $$\left[X_i,X_j\right] = -g_jZ_j\left(\ln g_i\right)X_i + g_iZ_i\left(\ln g_j\right)X_j+g_ig_j\left[Z_i,Z_j\right].$$
        From this formula the assumption follows.
       \qed
   \end{dow}
\end{tw}

\begin{uw}
    The example of the Lie module $\left\{|X_-,X_0,X_+|\right\}$ related to the Euler system discussed in Section \ref{euler-4.2} shows that the non-quasi-rectifiable Lie module can be rescaled to the quasi-rectifiable Lie algebra. Nevertheless, the basis of the Lie algebra obtained by rescaling is non-quasi-rectifiable. 
\end{uw}

The uniqueness of the rescalings for real Lie algebras is an essential step towards the proof of the uniqueness of rescalings for general Lie modules of vector fields. 

\begin{prop}\label{unik}
    Let $\left\{X_1,X_2,X_3\right\}$ be a real Lie algebra of vector fields with a fixed basis $\left\{X_1,X_2,X_3\right\}.$ 
    \begin{itemize}
        \item[$i)$] If the basis $\left\{X_1,X_2,X_3\right\}$ is quasi-rectifiable, then the real Lie algebra $\left\{X_1,X_2,X_3\right\}$ can be rescaled to a three-dimensional Abelian Lie algebra.
        \item[$ii)$] If the basis $\left\{X_1,X_2,X_3\right\}$ is non-quasi-rectifiable, then the real Lie algebra $\left\{X_1,X_2,X_3\right\}$ can be rescaled only to an isomorphic real Lie algebra.
    \end{itemize}
    \begin{dow}
        The statement of $i)$ is a direct consequence of Theorem \ref{modified}. We proceed to the proof of $ii).$
        Let us rewrite the commutation relations for the algebra $\left\{X_1,X_2,X_3\right\}$ and express the relations for the rescaled Lie algebra $\left\{Y_1,Y_2,Y_3\right\}$ in these terms. Let 
        \begin{equation}
            \begin{aligned}
                &\left[X_1,X_2\right]=a_1X_1+b_1X_2+c_1X_3,\quad
                \left[X_1,X_3\right]=a_2X_1+b_2X_2+c_2X_3,\\
                &\left[X_2,X_3\right]=a_3X_1+b_3X_2+c_3X_3.
            \end{aligned}
        \end{equation}
        The commutation relations of the vector fields $Y_1,Y_2,Y_3$ expressed by the above relations are
        \begin{equation}
            \begin{aligned}
                &\left[Y_1,Y_2\right] = g\left(a_1-X_2\left(\ln f\right)\right)Y_1 + f\left(b_1+X_1\left(\ln g\right)\right)Y_2+ c_1 \frac{fg}{h}Y_3,\\
                &\left[Y_1,Y_3\right] = h\left(a_2-X_3\left(\ln f\right)\right)Y_1 + b_2\frac{fh}{g}Y_2 + f\left(c_2+ X_1\left(\ln h\right)\right)Y_3,\\
                &\left[Y_2,Y_3\right] = a_3\frac{gh}{f}Y_1+ h\left(b_3-X_3\left(\ln g\right)\right)Y_2 + g\left(c_3+X_2\left(\ln h\right)\right)Y_3.
            \end{aligned}
        \end{equation}
        Let us denote the following structure constants of the Lie algebra by $t_1,t_2,t_3 \in \mathbb{R}:$
        $$t_3 := c_1\frac{fg}{h},\quad t_2 := \frac{fh}{g}b_2,\quad t_1:=\frac{gh}{f}a_3, \quad f g h \neq 0.$$
        The vanishing of some constant structure of $t_1,t_2,t_3$ is equivalent to the vanishing of the corresponding constant structure of the algebra $\left\{X_1,X_2,X_3\right\},$ that is $c_1,b_2,a_3$ and vice-versa. The rest of the structure constants of the rescaled Lie algebra $\left\{Y_1,Y_2,Y_3\right\}$ do not influence whether it is isomorphic (see the classification of three-dimensional Lie algebras given in \cite{Gru3}, Table 1, p. 39, or in \cite{Pat, Win, Win2}).
        This implies that the rescaled Lie algebra $\left\{Y_1,Y_2,Y_3\right\}$ is isomorphic to the Lie algebra $\left\{X_1,X_2,X_3\right\}.$
        \qed
    \end{dow}
\end{prop}

\begin{uw}
It is easy to see that the rescaling transformation of Lie modules and Lie algebras is transitive. That is, if $M_1, M_2, M_3 \in \mathcal{R}$ and the Lie module $M_2$ is obtained by a rescaling of $M_1,$ and the Lie module $M_3$ is obtained by a rescaling of $M_2,$ then the Lie module $M_1$ can be rescaled to the Lie module $M_3.$  
\end{uw}

\begin{tw}
Any $C^{\infty}$-Lie module of vector fields of the class $\mathfrak{M}_{\rm r}\left(N\right)$ can be rescaled to the unique (up to isomorphism) finite-dimensional real Lie algebra or the three-dimensional Abelian Lie algebra. 
\begin{dow}
This is an immediate consequence of Proposition \ref{unik} and the transitivity of the rescaling transform. \qed
\end{dow}
\end{tw}

\subsection{Wave superpositions in terms of Lie groups}

The following observation stresses that a quasi-rectifiable surface spanned by a pair of vector fields possesses the structure of a Lie group.

\begin{tw}
    Let $\left\{X_1,X_2,X_3\right\}$ be a real Lie algebra and let $M$ be a surface spanned by the vector fields $X_1$ and $X_2$ in the neighborhood of the point $p \in M.$ Let $\left\{X_1,X_2\right\}$ be a Lie subalgebra. Then locally, in the neighborhood of $p,$ the surface $M$ has the structure of a Lie group corresponding to the algebra $\left\{X_1,X_2\right\}.$
    \begin{dow}
        Let $\phi_1, \phi_2$ denote flows corresponding  to the vector fields $X_1$ and $X_2,$ respectively. Then the flows take the form $\phi_i(t) = \exp\left(t X_i\right),\; i \in \left\{1,2\right\}$ and any element in some fixed neighborhood of $p\in M$ can be expressed as $\exp\left(t_1 X_1\right)\exp\left(t_2 X_2\right)$ for some $t_1,t_2;$ for details, see \cite{Hel2}. In this way the unique simply-connected local Lie group is determined. 
        \qed
    \end{dow}
\end{tw}

\begin{uw}
    It is obvious that any vector fields which generate a Lie algebra corresponding to the Lie group $M$ are left-invariant with respect to $M.$
\end{uw}


\begin{tw}
If a Lie algebra $\left\{X_1,X_2,X_3\right\}$ takes the form $\left\{X_1,X_2,X_3\right\}=\left\{X_1,X_2\right\}\oplus \left\{X_3\right\},$ where the elements of the decomposition are Lie algebras, then the simply-connected Lie group $M_3$ corresponding to $\left\{X_1,X_2,X_3\right\}$ takes the form $M_3 = M_2 \times M_1,$ where $M_2,M_1$ are simply-connected Lie groups corresponding to $\left\{X_1,X_2\right\}$ and $\left\{X_3\right\},$ respectively. 
    \begin{dow}
        This observation follows from the general correspondence between Lie algebras and Lie groups.
        \qed
    \end{dow}
\end{tw}

Following \cite{Nom}, we adapt geometric notions to the setting under consideration; see also \cite{Hel2}.

\begin{defi}(K. Nomizu \cite{Nom})
Let $\mathcal{X}$ denote the set of all smooth vector fields on a smooth manifold $N$ and $\mathrm{End}\left(\mathcal{X}\right)$  denote the space of all endomorphisms of $\mathcal{X}.$
    An affine connection on $N$ is a mapping $t\colon \mathcal{X} \to \mathrm{End}\left(\mathcal{X}\right)$ satisfying the following two assumptions:
    \begin{itemize}
        \item[$i)$] $t\left(X+Y\right) = t\left(X\right)+t\left(Y\right),$
        \item[$ii)$] $t\left(fX\right)\left(Y\right) = ft\left(X\right)\left(Y\right) + \left(Yf\right)\left(X\right).$
    \end{itemize}
\end{defi}

The following proposition gives the correspondence between affine connections on the Lie group and bilinear forms.

\begin{tw}(K. Nomizu \cite{Nom})
    Let $G$ be a simply connected Lie group with corresponding Lie algebra $\mathfrak{g}.$ There exists a bijective mapping between the set of left-invariant affine connections on $G$ and the set of bilinear functions $b\colon \mathfrak{g}\times \mathfrak{g} \to \mathfrak{g}$ given by  $\left(t\left(X\right)\left(Y\right)\right)_e = b\left(Y,X\right)$ for any $X,Y \in \mathfrak{g}.$ \qed
\end{tw}

The next definition provides a more abstract notion of the parallel transport.

\begin{defi}(K. Nomizu \cite{Nom})
    Let $\phi_X\colon N\times I \to N$ be a flow of the vector field $X$ defined on the manifold $N.$ Let $Y$ be a vector field defined on $\phi_X\left(N\times I\right) \subset N.$ The parallel transport of the vector field $Y$ along the curve $\phi_X\left(N\times I\right)$ occurs if $t\left(Y\right)\left(X\right) =0$ at each point of $\phi_X\left(N\times I\right).$
\end{defi}

Among a class of affine connections on a given Lie group, it is possible to determine a unique one having affine properties.
The proof of the following theorem can be found in \cite{Nom}.

\begin{tw}(K. Nomizu \cite{Nom})$ $\\
    Let $G$ be a simply connected Lie group with corresponding Lie algebra $\mathfrak{g}.$ Let 
    \begin{equation}\label{nom1}
    t\left(Y\right)\left(X\right) := \frac{1}{2}\left[X,Y\right],\quad \text{ for any }\quad  X,Y  \in \mathfrak{g}.
    \end{equation}
    Then $t$ is a well-defined affine connection on the Lie group $G.$ Moreover, $t$ is both a left- and a right-invariant connection, and is the unique torsion-free connection whose $1$-parameter subgroups are geodesics. \qed
\end{tw}
\begin{uw}
    The affine connection \eqref{nom1} does not need to be a Levi-Civita connection on the Lie group $G.$ This is because, in the general case, a bi-invariant metric on the Lie group $G$ does not exist.
\end{uw}

The introduced notions allow us to provide a generalized counterpart of our result related to the Euler system involving a description of wave superpositions in terms of deformations of quasi-rectifiable surfaces.

\begin{tw}\label{groupdeco}
    Let the real Lie algebra $\left\{X,Y,Z\right\}$  take the form 
    \begin{equation}
    \left\{X,Y,Z\right\}=\left\{X,Y\right\}\oplus\left\{Z\right\}
    \end{equation}
    and let $M$ be a $1$-connected Lie group corresponding to $\left\{X,Y,Z\right\}.$ Then the Lie group $M$ is locally given as a parallel transport of surfaces spanned by $\left\{X,Y\right\}$ by the flow $\phi_Z$ of the vector field $Z$ in the affine connection \eqref{nom1}.
    \begin{dow}
        It is easy to see that $\left[X\oplus0,0\oplus Z\right]=\left[Y\oplus 0, 0\oplus Z\right] =0.$ This implies that $t\left(X\right)\left(Z\right)=t\left(Y\right)\left(Z\right)=0.$ Thus, both $X$ and $Y$ are transported in the parallel way along the flow $\phi_Z$ of the vector field $Z.$ Therefore the surface generated by $X,Y$ is under parallel transportation along $\phi_Z.$ 
        \qed
    \end{dow}
\end{tw}
\begin{uw}
    In Section \ref{ap6.3} of Appendix A, we apply the obtained results to a Lie algebra associated with waves corresponding to the Euler system. It should be remarked that the outcome obtained directly in the case of the Euler system in the previous section is slightly different. This is because we have chosen different systems of vector fields of the Lie module to avoid involved computations.
\end{uw}

We now discuss and summarize the results established in this section. The principal novelty is the generalization of the rescaling result of Section 5 to a certain class of abstract module of vector fields. The main advantage of this generalization lies in the possibility of reducing the essentially analytic structure of a module of vector fields to the purely algebraic framework of finite-dimensional real Lie algebras. This reduction enables the extraction of quantitative invariants that characterize a given family of vector fields.
	
    Proceeding in this manner, we identify these intrinsic properties of a given module of vector fields that determine whether it is quasi-rectifiable. The uniqueness statement proved in this section ensures that the correspondence between modules and Lie algebras induced by the rescaling transformation preserves all relevant structural information.
	
    We further advance the analysis of structures associated with a given module by assigning to each Lie algebra its corresponding Lie group. In addition, we prove a simple yet fundamental result (Theorem \ref{groupdeco}) which states  that if a Lie algebra of vector fields admits a decomposition into two Lie subalgebras of a specified type, then the associated Lie group can be described in terms of parallel transport. The significance of this theorem lies not merely in its analogy with the vector fields arising in the Euler system, but more importantly in the fact that it demonstrates that the parallel transport phenomenon observed in the study of the Euler system originates from the intrinsic structure of the rescaled Lie algebra.

\newpage

\noindent\textbf{\Large{Appendix A. On certain Lie module structures}}\\
\addcontentsline{toc}{section}{Appendix A. On certain Lie module structures}
\renewcommand{\thesection}{A}
\setcounter{section}{7}
\setcounter{equation}{0}
\setcounter{subsection}{0}
\setcounter{tw}{0}

\subsection{Structure of certain infinite-dimensional Lie algebras}

         \begin{tw}\label{witt_thm}
         The smallest (with respect to inclusion) real Lie algebra $\mathcal{L}$ containing the vector fields $X_+,X_0,X_-$ and their multiplications $\rho^{-n}X_+,\rho^{-n}X_0,\rho^{-n}X_-,\; n\in \left\{1,2,3,\dots\right\}$ defined on a three-dimensional manifold $N$ is isomorphic to the Virasoro algebra, which is the semidirect sum
        $$\mathcal{L} \simeq I_2 \roplus_{\Phi}\left(I_1 \roplus_{\Psi} W_{\geqslant -1}\right).$$ 
        \end{tw}
        \begin{dow}
            The smallest real Lie algebra with respect to inclusion containing the family\\ 
            $\left\{\rho^{-n}X_+,\rho^{-n}X_-,\rho^{-n}X_0\right\},\; n\in \left\{0,1,2,3,\dots\right\}$ and closed with respect to the Lie bracket is the real Lie algebra 
\begin{equation}\label{lie1}
\mathcal{L} =\spam \left\{X_+,X_-,X_0,\frac{1}{\rho}X_+,\frac{1}{\rho}X_-,\frac{1}{\rho}X_0,\frac{1}{\rho^2}X_+,\frac{1}{\rho^2}X_-,\frac{1}{\rho^2}X_0,\frac{1}{\rho^3}X_+,\dots\right\}.
\end{equation}
To make the presentation more clear we introduce the following notation
for $n, m, k \in \mathbb{N}\cup\{0\}$
\begin{equation}\label{lie2}
\begin{aligned}
a_n:=\frac{1}{\rho^n}X_+,\quad \; b_m:=\frac{1}{\rho^m}X_-,\quad c_k:=\frac{1}{\rho^k}X_0.
\end{aligned}
\end{equation}
The following bracket relations are derived from \eqref{eulercom}:

\begin{align*}\label{bra1}
&\left[a_n,b_m\right]_\mathcal{L} = \left(n-\frac{1}{2}\right) a_{n+m} + \left(\frac{1}{2}-m\right) b_{n+m},\\
&\left[a_n,c_k\right]_\mathcal{L} = \left(n+\frac{1}{4}\right) a_{n+k+1} - \frac{1}{4} b_{n+k+1} - \left(k+1\right)c_{n+k},\\
&\left[b_m,c_k\right]_\mathcal{L} = -\frac{1}{4} a_{m+k+1} + \left(m+\frac{1}{4}\right) b_{m+k+1} - \left(k+1\right) c_{n+k},\quad
\left[a_n, a_{n'}\right]_\mathcal{L} = \left(n-n'\right) a_{n+n'},\\
&\left[b_m,b_{m'}\right]_\mathcal{L} = \left(m-m'\right) b_{m+m'},\quad
\left[c_k,c_{k'}\right]_\mathcal{L} = \left(k-k'\right) c_{k+k'+1}.
\end{align*}

\noindent Thus, the  real Lie algebra under consideration takes the form 
\begin{equation}\label{lie3}
\mathcal{L} = \left(\left\{a_n,b_m,c_m\right\}_{n,m,k \in \mathbb{N}\cup \left\{0\right\}},\; \left[\cdot_1,\cdot_2\right]_\mathcal{L} \right).
\end{equation}

Let us choose a different basis of the Lie algebra $\mathcal{L}$ in the form $\left\{a_n,a_l-b_l,c_k\right\}_{n,m, l \in \mathbb{N}\cup \left\{0\right\}}$ and consider the subalgebra $\mathcal{L}\supset I_2=\left\{a_l-b_l\right\}_{l \in \mathbb{N}\cup \left\{0\right\}}$ of $\mathcal{L}$. It can easily be checked that $I_2$ is an ideal of the Lie algebra $\mathcal{L}$ and that  for  any $l, l' \in \mathbb{N}\cup \left\{0\right\}$ we have $\left[a_l-b_l,a_{l'}-b_{l'}\right]_I=0,$ thus 
\begin{equation}\label{iso1}
I_2 \simeq \left(\mathbb{R}[x], \left[\cdot_1,\cdot_2\right]_{I_2} \right), 
\end{equation}
in the sense of Lie algebras.

Let $W_{\geqslant -1}$ be the Witt algebra over $\mathbb{R}.$ We can identify $W_{\geqslant -1}$ with the Lie algebra of one-variable polynomials, that is $$W_{\geqslant -1} \simeq \left(\mathbb{R}[z], \left[\cdot_1,\cdot_2\right]_W\right),$$ where a bilinear bracket relation is given on the monomial basis $\left\{z^n\right\}_{n\in \mathbb{N}\cup \left\{0\right\}}$ by $$\left[z^n,z^m\right]_{W}=\left(n-m\right)z^{n+m}.$$
We independently introduce $I_1=\left(\mathbb{R}\left[y\right], \left[\cdot_1,\cdot_2\right]_{I_1} \equiv 0\right),$ i.e., a real algebra with the structure of an Abelian Lie algebra.
To obtain a relation between $I_1$ and $W_{\geqslant -1}$ we introduce a mapping ${\Psi\colon  W_{\geqslant -1} \to \der I_1}$ defined by the formula 
$$\Psi_{z^n}\left(y^m\right):= - \left(m+1\right) y^{m+n},\quad \Psi_0\left(y^m\right):= 0$$ on the  monomial basis. This formula can be uniquely extended by linearity to the whole space $W_{\geqslant -1}.$ For each polynomial $w\left(z\right) \in W_{\geqslant -1},$ we see that $\Psi_{w\left(z\right)} \in \der I_1,$ i.e., $\Psi_{w\left(z\right)}$ is a derivation of the ideal $I_1.$ 

We consider the semidirect sum $V:=I_1\roplus_{\Psi} W_{\geqslant -1}.$ Here,
$V$ can be naturally represented as 
$$V \simeq \left(\mathbb{R}\left[y\right]\oplus \mathbb{R}\left[z\right], \left[\cdot_1,\cdot_2\right]_{V}\right),$$ where the bracket $\left[\cdot_1,\cdot_2\right]_{V}$ is given by 
$$\left[y^{n_1}\oplus z^{m_1}, y^{n_2}\oplus z^{m_2}\right]_{V} = \left[y^{n_1},y^{n_2}\right]_{I_1} + \Psi_{z^{m_1}}\left(y^{n_2}\right) - \Psi_{z^{m_2}}\left(y^{n_1}\right) \oplus \left[z^{m_1},z^{m_2}\right]_{W}.$$
On the basis of the monomial terms, the above commutation relations take the form
\begin{equation}
\begin{aligned}
&\left[y^{n_1}\oplus0,0\oplus z^{m_2}\right]_{V} = -\left(n_1+1\right) y^{n_1+n_2}\oplus 0,\quad
\left[y^{n_1}\otimes 0, y^{n_2} \oplus 0\right]_{V}= 0 \oplus 0,\\
&\left[0 \oplus z^{m_1}, 0\oplus z^{m_2}\right]_{V} = 0\oplus \left[z^{m_1},z^{m_2}\right]_{W}.
\end{aligned}
\end{equation}

Now we return to our consideration of the algebra $\mathcal{L}$ and its ideal $I_2.$ The quotient Lie algebra $\mathcal{L}/I_2$ is defined as 
$$\mathcal{L}/I_2 := \left\{a_n+I_2,c_k+I_2\right\} = \left\{a_n+I_2, \left(c_k-a_{k+1}\right)+I_2\right\}= \left\{\tilde{a}_n,\tilde{c}_k\right\}=\left\{\tilde{a}_n, \tilde{c}_k-\tilde{a}_{k+1}\right\}$$ with the quotient Lie bracket 
\begin{equation}
\begin{aligned}
&\left[\tilde{a}_n,\tilde{c}_k\right]_/ := n\tilde{a}_{n+k+1} - \left(k+1\right)\tilde{c}_{n+k},\quad
&\left[\tilde{a}_n,\tilde{a}_{n'}\right]_/:= \left(n-n'\right)\tilde{a}_{n+n'},\\
&\left[\tilde{c}_k,\tilde{c}_{k'}\right]_/:=\left(k-k'\right) \tilde{c}_{k+k'+1}.
\end{aligned}
\end{equation}
In addition to the above, let us observe that 
\begin{equation}
\begin{aligned}
&\left[\tilde{a}_n,\tilde{c}_k-\tilde{a}_{k+1}\right]_/=-\left(k+1\right) \left(\tilde{c}_{n+k} - \tilde{a}_{n+k+1}\right),\quad
&\left[\tilde{c}_n,\tilde{c}_k-\tilde{a}_{k+1}\right]_/=-\left(k+1\right) \left(\tilde{c}_{n+k+1}-\tilde{a}_{n+k+2}\right),\\
&\left[\tilde{c}_n-\tilde{a}_{n+1},\tilde{c}_k-\tilde{a}_{k+1}\right]_/=0.
\end{aligned}
\end{equation}
Consequently, we have $V \simeq \mathcal{L}/I_2.$ Indeed, the homomorphism $Q_V\colon  \mathcal{L}/I_2 \to V$ given by the formula
\begin{equation}
\begin{aligned}
&Q_V\left(\tilde{a}_n\right) = z^n,\quad
Q_V\left(\tilde{c}_k-\tilde{a}_{k+1}\right) = y^k
\end{aligned}
\end{equation}
is the isomorphism of Lie algebras.

Further, we define a function $\Phi\colon V\to \der I_2,$ by the formula 
\begin{equation}
\begin{aligned}
&\Phi_{\tilde{a}_n}\left(a_l-b_l\right):=\left(\frac{1}{2}-l\right) \left(a_{n+l}-b_{n+l}\right),\quad
\Phi_{\tilde{c}_k}\left(a_l-b_l\right):=-\left(l+\frac{1}{2}\right)\left(a_{l+k+1}-b_{l+k+1}\right).
\end{aligned}
\end{equation}
By the isomorphism \eqref{iso1}, the mapping $\Phi$ can be represented as 
\begin{equation}
\begin{aligned}
&\Phi_{\tilde{a}_n}\left(x^l\right) = \left(\frac{1}{2}-l\right) x^{n+l},\quad
\Phi_{\tilde{c}_k}\left(x^l\right) = -\left(l+\frac{1}{2}\right)x^{l+k+1}.
\end{aligned}
\end{equation}
The semidirect sum $I_2\roplus_{\Phi} W_{\geqslant -1}$ which is a set $\mathbb{R}\left[x\right]\oplus V$ is equipped with a Lie bracket $\left[\cdot_1,\cdot_2\right]_{\Phi}$ given on the basis by
\begin{equation}
\begin{aligned}
&\left[x^{k_1}\oplus\left(0\oplus 0\right),0 \oplus \left(y^{n_1}\oplus 0\right)\right]_{\Phi} = \left(k_1+\frac{1}{2}\right) x^{k_1+n_1+1}\oplus \left(0\oplus 0\right),\\
&\left[x^{k_1}\oplus \left(0 \oplus 0\right),0 \oplus \left(0\oplus z^{m_1}\right)\right]_{\Phi} = \left(k_1-\frac{1}{2}\right) x^{k_1+m_1}\oplus \left(0\oplus 0\right).
\end{aligned}
\end{equation}
From these relations, it follows that $I_2\roplus_{\Phi} V \simeq \mathcal{L}.$ 
Denoting $\left[\cdot_1,\cdot_2\right]_{\mathcal{A}}:=\left[\cdot_1,\cdot_2\right]_{\Phi},$ we see that 
$$\mathcal{L} \simeq \mathcal{A} = \left(I_2\roplus_{\Phi} V,\left[\cdot_1,\cdot_2\right]_{V}\right) = \left(I_2 \roplus_{\Phi}\left(I_1\roplus_{\Psi}W_{\geqslant -1}\right),[\cdot_1,\cdot_2]_{V}\right)= 
\left(\mathbb{R}\left[x\right]\oplus \left(\mathbb{R}\left[y\right]\oplus \mathbb{R}\left[z\right]\right),\left[\cdot_1,\cdot_2\right]_{\mathcal{A}}\right).$$
In this way we have established the representations of the infinite-dimensional real Lie algebras, namely the Virasoro algebra $\mathcal{L}$ and the Witt algebra. \qed
        \end{dow}

In an analogous way, a characterization of the Lie subalgebra generated by sequences of $\rho^{-n}X_-, \rho^{-n}X_+$ can be given. Since the proof does not produce further difficulties and follows the lines of the proof of the above theorem, we omit it. 
        
         \begin{tw}\label{witt_thm2}
         The smallest (with respect to inclusion) real Lie algebra $\mathcal{H}$ containing two vector fields $X_+, X_-$ and their consecutive multiplications $a_n=\rho^{-n}X_+,\; b_n=\rho^{-n} X_-,\; n\in \mathbb{N}$ is isomorphic to
        $$\mathcal{H} \simeq I_3\roplus_{\mu}W_{\geqslant -1},$$
        where $I_3$ is an Abelian Lie algebra, $\mu$ is a shift operator, and $W_{\geqslant -1}$ stands for the Witt algebra over $\mathbb{R}.$ \qed
        \end{tw}

The above theorems describe structures of all iterations of waves, i.e., $\rho^nX_+,\; \rho^nX_-,\; \rho^nX_0$. Thus, informally speaking, they describe the `free-structure' of iterations of waves that appear in the analysis of the Euler system.

\subsection{Rescaling of the $C^\infty$-Lie module to a non-quasi-rectifiable Lie algebra}\label{ap6.1}

A classification of real Lie algebras in $\mathbb{R}$ from the point of view of quasi-rectifiability was performed in \cite{Gru3}. This classification can be of use for establishing the quasi-rectifiability of Lie modules of particular PDE systems as illustrated below.
\begin{exam}
    Let us consider the following linear system in the evolutionary form 
    \begin{equation}\label{vectheis}
       \frac{\partial}{\partial t}u + A \frac{\partial}{\partial x}u=0, 
    \end{equation}
    where $$A = \begin{pmatrix}
        \lambda_1&0&0\\
        0&1&0\\
        0&x\left(1-\lambda_3\right)&\lambda_3
    \end{pmatrix},$$
    for some smooth functions $\lambda_1, \lambda_3$ of $x$, independent of the variable $t.$ 
    It is easy to see that for the eigenvalues $\lambda_1, 1$ and $\lambda_3$ the particular corresponding eigenvectors can be chosen to be $$X_1 = \left(1,0,0\right),\quad X_2=\left(0,x,x^2\right),\quad X_3 = \left(0,0,1\right),$$ respectively. 
    Up to antisymmetry, the Lie brackets of pairs of the above vector fields are given by
    \begin{equation}
        \begin{aligned}
            \left[X_1,X_2\right] = \frac{1}{x}X_2 + xX_3,\quad
            \left[X_1,X_3\right] = 0,\quad
            \left[X_2,X_3\right]=0.
        \end{aligned}
    \end{equation}

    Let $Y_1 = fX_1,\; Y_2 = gX_2,\; Y_3 = hX_3,$ where the non-zero scaling functions $f,g,h \in C^{\infty}$ of the variables $x,y,z$ are going to be determined in further computations.

   The structure constants $c_i,\; i\in \left\{1,\dots,9\right\},$ enumerated according to \eqref{stale}, of the rescaled Lie algebra $\left\{Y_1,Y_2,Y_3\right\}$ can be computed as follows 
    \begin{equation}
        \begin{aligned}
            \left[Y_1,Y_2\right] = - \frac{g}{f}X_2\left(f\right)Y_1+\frac{1}{g}\left(fg\frac{1}{x}+fX_1\left(g\right)\right)Y_2+\frac{fg}{h}xY_3,
        \end{aligned}
    \end{equation}
    and thus 
    \begin{equation}
    c_1=-\frac{g}{f}X_2\left(f\right),\quad c_2=\frac{1}{g}\left(fg\frac{1}{x}+fX_1\left(g\right)\right),\quad c_3=\frac{fg}{h}x.
    \end{equation}
    By rewriting the other two Lie brackets 
    \begin{equation}
        \begin{aligned}
            &\left[Y_1,Y_3\right]=-\frac{h}{f}X_3\left(f\right)Y_1+\frac{f}{h}X_1\left(h\right)Y_3,\\
            &\left[Y_2,Y_3\right] = - \frac{h}{g}X_3\left(g\right)Y_2+\frac{g}{h}X_2\left(h\right)Y_3,
        \end{aligned}
    \end{equation}
    and using the form of the vector fields $X_1, X_2, X_3,$ we obtain
    \begin{equation}\label{strcomp}
        \begin{aligned}
            &c_1 = -g\left(x\frac{\partial \ln f}{\partial y}+x^2\frac{\partial \ln f}{\partial z}\right),\quad
            c_2 = f\left(\frac{1}{x}+ \frac{\partial \ln g}{\partial x}\right),\quad
            c_3 = \frac{fg}{h}x,\quad
            c_4 = -h \frac{\partial \ln f}{\partial z},\\
            &c_5 = 0,\quad
            c_6 = f \frac{\partial \ln h}{\partial x},\quad
            c_7 = 0,\quad
            c_8 = -h\frac{\partial \ln g}{\partial z},\quad
            c_9 = g\left(x \frac{\partial \ln h}{\partial y}+ x^2\frac{\partial \ln h}{\partial z}\right).
        \end{aligned}
    \end{equation}
    The fact that we look for nonvanishing functions $f,g,h$ immediately implies that $c_3 \neq 0.$ Thus, we may take $$h = \frac{fg}{c_3}x.$$ Applying this to the equation for the constant $c_6$, we obtain 
    \begin{equation}
        \begin{aligned}
            c_6 = f \frac{\partial \ln f}{\partial x} + f\left(\frac{\partial \ln g}{\partial x}+ \frac{1}{x}\right).
        \end{aligned}
    \end{equation}
By the formula for the   constant $c_2$, we obtain
\begin{equation}\label{c6}
c_6 = \frac{\partial f}{\partial x}+c_2,
\end{equation}
which implies that 
\begin{equation}\label{funkc}
f = \left(c_6-c_2\right)x + c\left(y,z\right).
\end{equation}
The function $c=c\left(y,z\right)$ is a priori dependent on the variables $y$ and $z.$ It can be shown directly that $c$ needs to be constant. As we aim to construct only one possible transformation, we can take this as an assumption. By the expression determining $c_6,$ under the assumption that $c_2\neq c_6,$ we have $$\frac{c_6}{\left(\left(c_6-c_2\right)x+c\right)}=\frac{1}{h}\frac{\partial h}{\partial x}.$$ 
This implies that, for a certain nonvanishing function $\widetilde{c}=\widetilde{c}(y,z),$ we have
$$h = e^{\widetilde{c}}\left(\left(c_6-c_2\right)x+c\right)^{\frac{c_6}{c_6-c_2}},$$ and
$$g = \frac{c_3h}{\left(c_6-c_2\right)x^2+cx}=\frac{c_3 e^{\widetilde{c}}}{x}\left(\left(c_6-c_2\right)x+c\right)^{\frac{c_2}{c_6-c_2}}.$$ From \eqref{strcomp}, we obtain that 
$$c_1=c_4=c_5=c_7=c_8=c_9=0.$$
The commutation relations of the Lie algebra $\left\{Y_1,Y_2,Y_3\right\}$ take the following form:
\begin{equation}
    \begin{aligned}
        \left[Y_1,Y_2\right]=c_2Y_2+c_3Y_3,\quad
        \left[Y_1,Y_3\right]=c_6Y_3,\quad
        \left[Y_2,Y_3\right]=0.
    \end{aligned}
\end{equation}
Now, let us insert the obtained formulas for the scaling functions $f,g$ and $h$ into an expression determining the nonvanishing structure constants $c_2,c_3,c_6.$ Considering the expression for $c_2$ and inserting the function $g,$ we obtain 
$$c_2 = \frac{c_2}{c_6-c_2}\left(\left(c_6-c_2\right)x+c\right)^{-\frac{c_2}{c_6-c_2}}.$$ Further, this gives $$\left(-\frac{c_6-c_2}{c_2}\right)\ln \left(c_6-c_2\right)=\ln \left(\left(c_6-c_2\right)x+c\right).$$ As the left-hand side is $x$-independent we conclude that $c_6=c_2,$ which is a contradiction.
Thus, $c_2=c_6$ and by \eqref{funkc} we have $$f=c,$$ so the nonvanishing assumption implies that $c\neq 0.$ Moreover $$h = \frac{1}{c_3}cxg,$$ and by the $c_6$-equation $$\ln h = \frac{c_6}{c}x+\widetilde{c},\quad \text{ hence }\quad
 h=e^{\widetilde{c}}e^{\frac{c_6}{c}x},$$ and 
$$g = \frac{c_3}{c}\frac{e^{\widetilde{c}}e^{\frac{c_6}{c}x}}{x}.$$
For simplicity, let us assume that the function $\widetilde{c}$ is constant. 
Since $$c_2 = c\left(\frac{1}{x}+ \frac{\partial \ln g}{\partial x}\right)\quad \text{ and }\quad \frac{\partial \ln g}{\partial x}=\left(\frac{c_6}{c}-\frac{1}{cx}\right),$$  it follows that $$c_2=c_6=0,\quad \text{ and }\quad c=1.$$
Let us put $c':=e^{\widetilde{c}}.$
Then we have
\begin{equation}
    \begin{aligned}
        f=1,\quad
        h=c',\quad
        g=c_3c'\frac{1}{x}.
    \end{aligned}
\end{equation}
Consequently, the Lie algebra $\left\{Y_1,Y_2,Y_3\right\}$ takes the form
\begin{equation}
    \begin{aligned}
        \left[Y_1,Y_2\right]=c_3Y_3,\quad
        \left[Y_1,Y_3\right]=0,\quad
        \left[Y_2,Y_3\right]=0.
    \end{aligned}
\end{equation}
We see that, as $c_3\neq 0,$ this algebra is isomorphic to the three-dimensional Heisenberg algebra which was established in \cite{Gru3} to be not quasi-rectifiable. This means that there does not exist a basis of the Lie module $\left\{|X_1,X_2,X_3|\right\}$ which is quasi-rectifiable. This implies that there does not exist a parametrization of a manifold of wave superpositions corresponding to system \eqref{vectheis} of the form presented in Section \ref{7.1}.
\end{exam}

\subsection{Certain properties of the rescaling transform}
The rescaling transform is studied here from the perspective of a mapping between the class $\mathfrak{M}_{\rm r}$ of Lie modules and a class of isomorphism classes of real Lie algebras. 

\begin{defi}\label{resc2}
    Let $\mathrm{Lie}_{\mathbb{R}}/_{\sim}$ denote a class of isomorphism classes of three-dimensional real Lie algebras and let $\mathfrak{M}_{\rm r}^B:=\left\{\left(R,B\right)\colon \; R\in\mathfrak{M}_{\rm r},\; B\text{ is a basis of } R\right\}.$ Let us define the mapping $\mathcal{F}\colon  \mathfrak{M}_{\rm r}^B \to \mathrm{Lie}_{\mathbb{R}}/_{\sim}$ such that for any pair $\left(M,B\right)\in \mathfrak{M}_{\rm r}^B$ it prescribes the class of isomorphic real Lie algebras obtained by a rescaling transformation or the Abelian Lie algebra.  
\end{defi}

\begin{prop}
    The mapping $\mathcal{F}\colon  \mathfrak{M}_{\rm r}^B \to \mathrm{Lie}_{\mathbb{R}}/_{\sim}$ is well-defined.
    \begin{dow}
        For any Lie module $M \in \mathfrak{M}_{\rm r}$ with a chosen basis $B\subset M$ we can prescribe the class of isomorphic Lie algebras obtained by a rescaling transformation. By the uniqueness result (Theorem 6.14), for each $\left(M,B\right) \in \mathfrak{M}_{\rm r}^B$ there exists either exactly one such class or an Abelian Lie algebra. This implies that $\mathcal{F}$ is a well-defined mapping. 
        \qed
    \end{dow}
\end{prop}

\begin{uw}
It is interesting to examine functorial properties of the mapping $\mathcal{F}$ between relevant categories. We leave this question for further studies. 
\end{uw}

\subsection{Lie group associated with the Euler system}\label{ap6.3}
The transform $\mathcal{F}$ (Def. \ref{resc2})  applied to the $C^{\infty}$-Lie module $\left\{|X_-,X_0,X_+|\right\}$ of waves related to the Euler system \eqref{euler} gives the real Lie algebra $$\mathcal{F}\left(\left(\left\{|X_-,X_0,X_+|\right\},\left\{X_-,X_0,X_+\right\}\right)\right) = \left\{Y_0,Y_1,Y_2\right\}$$ of the form
\begin{equation}
    \begin{aligned}
        \left[Y_1,Y_2\right] =0,\quad
        \left[Y_0,Y_2\right]=Y_2,\quad
        \left[Y_0,Y_1\right]=0.
    \end{aligned}
\end{equation}
In the rescaled basis we can explicitly express this as $$Y_0 = \frac{1}{2\beta}\bar{Y}_0,\quad Y_1= \beta \bar{Y}_+ - \alpha\bar{Y}_-+ \frac{r}{\beta}\bar{Y}_0,\quad Y_2 = \alpha \bar{Y}_- + \beta \bar{Y}_+,$$
for certain real constants $\alpha, \beta, r.$

The real Lie algebra $\left\{Y_0,Y_1,Y_2\right\}$ is isomorphic to $L\left(2,1\right)\oplus L\left(1,0\right),$ where $L\left(1,0\right)$ is the unique one-dimensional Lie algebra and $L\left(2,1\right)$ is the unique non-abelian two-dimensional Lie algebra. Indeed, let $L\left(2,1\right) = \left\{e_1,e_2\right\}$ with the Lie bracket relation $\left[e_1,e_2\right]=e_1,$ and $L\left(1,0\right)=\left\{e_3\right\}.$ Then the mapping $\Psi\colon  \left\{Y_0,Y_1,Y_2\right\} \to L\left(2,1\right) \oplus L\left(1,0\right)$ given by 
$$\Psi\left(-Y_0\right) = e_2\oplus0,\quad \Psi\left(Y_2\right) = e_1\oplus 0,\quad \Psi\left(Y_1\right)=0\oplus e_3$$ is an isomorphism of real Lie algebras.

A Lie group of the algebra $L\left(1,0\right)$ is $\left(\mathbb{R},+\right)$ and a Lie group of $L\left(2,1\right)$ is the group of affine transformations, which can be realized as $\left(\mathbb{R}^2,\ast\right),$ where $\left(c,d\right)\ast \left(c',d'\right) = \left(c+c',d+e^cd'\right).$ We know that the simply-connected Lie group corresponding to $L\left(2,1\right)\oplus L\left(1,0\right)$ takes the form $\left(\mathbb{R}^2,\ast\right)\times\left(\mathbb{R},+\right).$ 

For clarity, let us denote $$\tau_1 = \bar{Y}_- + \bar{Y}_+,\quad \tau_2 = -\frac{1}{2\beta}\bar{Y}_0,\quad \tau_3 = \beta \bar{Y}_+ - \alpha \bar{Y}_- + \frac{r}{\beta}\bar{Y}_0.$$ The above observations show that the manifold of superpositions of waves is diffeomorphic to the product $M_{\tau_1,\tau_2}\times M_{\tau_3},$ where $M_{\tau_3}$ is the $1$-dimensional manifold diffeomorphic to the vector field $\tau_3$ and $M_{\tau_1,\tau_2}$ is the surface diffeomorphic to the surface spanned by $\tau_1,\tau_2.$ Besides that, the diffeomorphisms induce the isomorphisms of corresponding Lie algebras. Therefore, these diffeomorphisms are isomorphisms in the sense of Lie groups.\\

\noindent\textbf{\Large{Appendix B. A class of explicit solutions of the reduced Euler system}}\\
\addcontentsline{toc}{section}{Appendix B. A class of explicit solutions of the reduced Euler system}
\renewcommand{\thesection}{B}
\setcounter{section}{8}
\setcounter{equation}{0}
A particular class of non-elastic solutions of the reduced Euler system \eqref{reduce} can be found if we assume that the function $t_3$ is expressible in  the separable form 

\begin{equation}
    t_3=A\left(x\right)+B\left(t\right) \label{separable 1}.
\end{equation}
Consequently, from the third equation of \eqref{reduce}, we obtain 
\begin{equation}
    t_2= \frac{{B}'\left(t\right)}{2\sqrt{3} {A}'\left(x\right)},
    \label{separable 2}
\end{equation}
where the prime $'$ is introduced to denote interchangeably a shortening of notation of the derivative with respect to the argument of functions $A$ and $B,$ respectively. Substituting   \eqref{separable 1} and \eqref{separable 2} into the second equation of \eqref{reduce} we determine that $\frac{\partial t_1}{\partial x}$ takes the form

\begin{equation}
    \frac{\partial t_1}{\partial x}= \frac{e^ {-4 t_1 +A\left(x\right)+B\left(t\right)}}{6\left({A}'\left(x\right)\right)^3} \left( B'\left(t\right)^2 {A}''\left(x\right) +{A}'\left(x\right)^2 {B}''\left(t\right) \right).
    \label{separate 3}
\end{equation}

\noindent Eliminating $\frac{\partial t_1}{\partial x}$ in the first equation of \eqref{reduce}, we find

\begin{equation}
    \frac{\partial t_1}{\partial t}= \frac{e ^{-4 t_1}}{6 \left({A}'\right)^4} \left( e^{A\left(x\right)+B\left(t\right)}\left({B}'\right)^3 {A}'+{A}'^2 {B}' \left( -3 e^{4 t_1} {A}''+e^{A\left(x\right)+B\left(t\right)} {B}''(t) \right) \right).
    \label{separate 4}
\end{equation}

\noindent The compatibility condition for $\frac{\partial t_1}{\partial x}$ and $\frac{\partial t_1}{\partial t}$ allows us to find two different real classes of solutions for  $t_1:$

\begin{equation}\label{nonelastic3}
\begin{aligned}
& t_1= {\frac{1}{4} \Bigg(-\log} (-(A'(x)^2 (-6 A''\left(x\right)^2 B'\left(t\right)^4+6 A'\left(x\right)^6 B''\left(t\right)-3 A'\left(x\right)^4 B''\left(t\right)^2\\
&-3 B^{\left(3\right)}\left(t\right) A'\left(x\right)^4 B'\left(t\right)  
 -6A'\left(x\right)^4 B'\left(t\right)^2 B''\left(t\right)+3 A^{\left(3\right)}\left(x\right) A'\left(x\right)^3 B'\left(t\right)^2\\
 &-3 A'\left(x\right)^4 A''\left(x\right) B''\left(t\right)+6 A'\left(x\right)^4 A''\left(x\right) B'\left(t\right)^2
 -9 A'\left(x\right)^2 A''\left(x\right)^2 B'\left(t\right)^2\\
 &-6 A'\left(x\right)^2 A''\left(x\right) B'\left(t\right)^4 \\
 &-15 A'\left(x\right)^2 A''\left(x\right) B'\left(t\right)^2 B''\left(t\right))) -(A'\left(x\right)^4 (-6 A''\left(x\right)^2 B'\left(t\right)^4+6 A'\left(x\right)^6 B''\left(t\right)\\
&-3 A'\left(x\right)^4 B''\left(t\right)^2-3 B^{\left(3\right)}\left(t\right) A'\left(x\right)^4 B'\left(t\right)  -6 A'\left(x\right)^4 B'\left(t\right)^2 B''\left(t\right) +3 A^{\left(3\right)}\left(x\right) A'\left(x\right)^3 B'\left(t\right)^2\\
&-3 A'\left(x\right)^4 A''\left(x\right) B''\left(t\right) +6 A'\left(x\right)^4 A''\left(x\right) B'\left(t\right)^2
-9 A'\left(x\right)^2 A''\left(x\right)^2 B'\left(t\right)^2\\
&-6 A'\left(x\right)^2 A''\left(x\right) B'\left(t\right)^4 
-15 A'\left(x\right)^2 A''\left(x\right) B'\left(t\right)^2 B''\left(t\right))^2\\
 &-36 A'\left(x\right)^6 A''\left(x\right) B''\left(t\right) 
 (2 A''\left(x\right)^2 B'\left(t\right)^6 -2 A'\left(x\right)^6 B''\left(t\right)^2\\
 &+2 A'\left(x\right)^4 B'\left(t\right)^2 B''\left(t\right)^2 -2 A'\left(x\right)^2 A''\left(x\right)^2 B'\left(t\right)^4 -4 A'\left(x\right)^4 A''\left(x\right) B'\left(t\right)^2 B''\left(t\right)\\
 &+4 A'\left(x\right)^2 A''\left(x\right) B'\left(t\right)^4 B''\left(t\right)))^{\frac{1}{2}})
 (2 (2 A''\left(x\right)^2 B'\left(t\right)^6-2 A'\left(x\right)^6 B''\left(t\right)^2\\
 &+2 A'\left(x\right)^4 B'\left(t\right)^2 B''\left(t\right)^2-2 A'\left(x\right)^2 A''\left(x\right)^2 B'\left(t\right)^4 -4 A'\left(x\right)^4 A''\left(x\right) B'\left(t\right)^2 B''\left(t\right) \\
 &+4 A'\left(x\right)^2 A''\left(x\right) B'\left(t\right)^4 B''\left(t\right)))^{-1})+A\left(x\right){+B(t)\Bigg),}
\end{aligned}
\end{equation}
and 
\begin{equation}\label{nonelastic4}
\begin{aligned}
& t_1= \frac{1}{4} {\bigl(}-\log {\bigl(}{\bigl(}A'(x)^4 {\bigl(}-6 A''(x)^2 B'(t)^4+6 A'(x)^6 B''(t)-3 A'(x)^4 B''(t)^2\\&-3 B^{(3)}(t) A'(x)^4 B'(t)
 -6A'(x)^4 B'(t)^2 B''(t)+3 A^{(3)}(x) A'(x)^3 B'(t)^2-3 A'(x)^4 A''(x) B''(t)\\&+6 A'(x)^4 A''(x) B'(t)^2-9 A'(x)^2 A''(x)^2 B'(t)^2-6 A'(x)^2 A''(x) B'(t)^4\\&-15 A'(x)^2 A''(x) B'(t)^2 B''(t){\bigr)}^2 -36 A'(x)^6 A''(x) B''(t) {\bigl(}2 A''(x)^2 B'(t)^6-2 A'(x)^6 B''(t)^2\\&+2 A'(x)^4 B'(t)^2 B''(t)^2 -2 A'(x)^2 A''(x)^2 B'(t)^4 -4 A'(x)^4 A''(x) B'(t)^2 B''(t)\\&+4 A'(x)^2 A''(x) B'(t)^4 B''(t){\bigr)}{\bigr)}^{\frac{1}{2}} -A'(x)^2 {\bigl(}-6 A''(x)^2 B'(t)^4+6 A'(x)^6 B''(t)\\&-3 A'(x)^4 B''(t)^2-3 B^{(3)}(t) A'(x)^4 B'(t) -6 A'(x)^4 B'(t)^2 B''(t)+3 A^{(3)}(x) A'(x)^3 B'(t)^2\\&-3 A'(x)^4 A''(x) B''(t)+6 A'(x)^4 A''(x) B'(t)^2 -9 A'(x)^2 A''(x)^2 B'(t)^2-6 A'(x)^2 A''(x) B'(t)^4\\&-15 A'(x)^2 A''(x) B'(t)^2 B''(t){\bigr)} {\bigl(}2 {\bigl(}2 A''(x)^2 B'(t)^6-2 A'(x)^6 B''(t)^2+2 A'(x)^4 B'(t)^2 B''(t)^2\\&-2 A'(x)^2 A''(x)^2 B'(t)^4 -4 A'(x)^4 A''(x) B'(t)^2 B''(t)+4 A'(x)^2 A''(x) B'(t)^4 B''(t){\bigr)}{\bigr)}^{-1}{\bigr)}\\&+A(x)+B(t){\bigr)}.
\end{aligned}
\end{equation}

\noindent The non-elastic wave superposition of the entropic wave E with one of the sound waves $S_+$ or $S_-$ is described by the expressions \eqref{param} 

\begin{equation}
    \rho=e^{2t_1+t_3}, \quad p=e^{6t_1}, \quad u=2\sqrt{3}t_2, \nonumber
\end{equation}
where the function $t_1$ is given by one of the expressions \eqref{nonelastic3} and \eqref{nonelastic4} and the functions $t_2$ and $t_3$ are given by \eqref{separable 2} and \eqref{separable 1}, respectively. This results in the production of slow sound waves $S_+$ or $S_-$ in the region of interaction $v(\mathcal{M})$ (i.e. the non-elastic wave superposition $\mathfrak{X}_\mathcal{M}=1$). The interaction can be denoted by $S_\pm+E \xrightarrow{}S_++ E+S_-. $ \\

\noindent\textbf{\Large{Acknowledgments}}\\

\addcontentsline{toc}{section}{Acknowledgments}
\setcounter{section}{9}
\setcounter{equation}{0}
\noindent A. M. Grundland was partially supported by an Individual Research Grant from NSERC of Canada. Ł. Chomienia would like to thank the Centr{e} de Recherches Math{\'e}matiques (CRM), Universit{\'e} de Montr{\'e}al for its warm hospitality during his stay as a postdoctoral fellow and also acknowledges partial financial support provided by the Mathematical Physics Laboratory of the CRM, Universit{\'e} de Montr{\'e}al. This work was partially sponsored by the Partenariat Acad{\'e}mique between the CRM and the Université du Québec à Trois-Rivières.\\\\\\

\noindent\textbf{\Large{Declarations}}\\

\addcontentsline{toc}{section}{Declarations}
\setcounter{section}{9}
\setcounter{equation}{0}
\noindent Competing Interests: The authors have no relevant financial or non-financial interest to disclose and no competing interests to declare that are relevant to the content of this article.\\\\
Authors' Contributions: Each of the two authors made a 50\% contribution to the work.\\\\
Funding: National Science and Engineering Research Council (NSERC) of Canada and Mathematical Physics Laboratory of the Centre de Recherches Mathématiques (CRM).\\\\

\end{document}